\documentclass[final,3p,times,onecolumn]{elsarticle}

\newcommand{\largeSinglePlotSize}{0.90\columnwidth}
\newcommand{\singlePlotSize}{0.80\columnwidth}
\newcommand{\doublePlotSize}{0.45\columnwidth}
\journal{Physical Review D}

\usepackage{etoolbox}
\newtoggle{isPRD}
\togglefalse{isPRD}

\usepackage{graphicx}
\usepackage{dcolumn}
\usepackage{bm}
\usepackage{atlasphysics} 
\usepackage{multirow}
\usepackage{amsmath}
\usepackage{float}
\usepackage{amssymb} 
\usepackage{subfigure}
\usepackage{rotating}
\usepackage{wasysym}
\usepackage{xspace}

\newcommand{\atlasnote}[1]{\def\@atlasnote{#1}}

\newcommand{\ptt}{\ensuremath{p_{\mathrm{Tt}}}}

\newcommand{\hgg}{\ensuremath{H\rightarrow \gamma\gamma}}

\newcommand{\hZZllll}{\ensuremath{H{\rightarrow\,}ZZ^{*}{\rightarrow\,}4\ell}}

\newcommand{\hzzllll}{\ensuremath{H \rightarrow ZZ^{*}\rightarrow 4\ell}}

\newcommand{\mh}{\ensuremath{m_{H}}}

\newcommand{\ztollg}{\ensuremath{Z{\rightarrow\,}\ell^+\ell^-\gamma}}
\newcommand{\ztoee}{\ensuremath{Z{\rightarrow\,}e^+e^-}}
\newcommand{\Jpsitoee}{\ensuremath{J/\psi{\rightarrow\,}e^+e^-}}

\newcommand{\ZZbkg}{\ensuremath{ZZ^{*}}}

\mathchardef\mhyphen="2D

\def\Zmm{{Z \to \mu^{+} \mu^{-}}}
\def\Jpsimm{{J/\psi \to \mu^{+} \mu^{-}}}

\def\vec#1{{\mbox{$\boldsymbol{#1}$}}}

\usepackage{amssymb}

\usepackage{lineno}
\usepackage{chngpage}

\usepackage{hypernat}

\usepackage{hyperref}

\newcommand{\papertitle}{Measurement of the Higgs boson mass from the \hgg\ and \hZZllll\ channels with the ATLAS detector using 25 \ifb\ of $pp$ collision data} 

\usepackage{preprintcover}  
\PreprintCoverPaperTitle{\papertitle}  
\PreprintIdNumber{CERN-PH-EP-2014-122}  
\PreprintCoverAbstract{
  An improved measurement of the mass of the Higgs boson is derived from a combined fit to the
  invariant mass spectra of the decay channels \hgg\ and \hZZllll. The analysis uses the $pp$
  collision data sample recorded by the ATLAS experiment at the CERN Large Hadron Collider at
  center-of-mass energies of 7 TeV and 8~TeV, corresponding to an integrated luminosity of 25\,\ifb.
  The measured value of the Higgs boson mass is $m_{H} = 125.36 \pm 0.37 \,\mathrm{(stat)} \pm 0.18
  \,\mathrm{(syst)} \gev$.  This result is based on improved energy-scale calibrations for photons,
  electrons, and muons as well as other analysis improvements, and supersedes the previous result
  from ATLAS. Upper limits on the total width of the Higgs boson are derived from fits to the
  invariant mass spectra of the \hgg\ and \hZZllll\ decay channels.
  }  
\PreprintJournalName{Physical Review D}  

\begin{document}

\title{\bf Measurement of the Higgs boson mass from the \hgg\ and \hZZllll\ channels with the ATLAS detector using 25 \ifb\ of $pp$ collision data}

\begin{abstract}

  An improved measurement of the mass of the Higgs boson is derived from a combined fit to the
  invariant mass spectra of the decay channels \hgg\ and \hZZllll. The analysis uses the $pp$
  collision data sample recorded by the ATLAS experiment at the CERN Large Hadron Collider at
  center-of-mass energies of 7 TeV and 8~TeV, corresponding to an integrated luminosity of 25\,\ifb.
  The measured value of the Higgs boson mass is $m_{H} = 125.36 \pm 0.37 \,\mathrm{(stat)} \pm 0.18
  \,\mathrm{(syst)} \gev$.  This result is based on improved energy-scale calibrations for photons,
  electrons, and muons as well as other analysis improvements, and supersedes the previous result
  from ATLAS. Upper limits on the total width of the Higgs boson are derived from fits to the
  invariant mass spectra of the \hgg\ and \hZZllll\ decay channels.
\end{abstract}

\maketitle


\hyphenation{ATLAS}

\section{Introduction\label{sec:introduction}}
In 2012, the ATLAS and CMS Collaborations published the discovery of a new
particle~\cite{HiggsObservationATLAS, HiggsObservationCMS} in the search for the Standard Model (SM)
Higgs boson~\cite{Englert:1964et,Higgs:1964ia,Higgs:1964pj,Guralnik:1964eu,Higgs:1966ev,Kibble:1967sv} at
the CERN Large Hadron Collider (LHC)~\cite{1748-0221-3-08-S08001}.  In the SM, the Higgs boson mass
is not predicted. Its measurement is therefore required for precise calculations of electroweak
observables including the production and decay properties of the Higgs boson itself.  These
calculations are needed to test the coupling structure of the SM Higgs boson, as suggested in
Ref.~\cite{Dittmaier:2013} and references therein.

The LHC Collaborations have chosen a model-independent approach to measure the Higgs boson mass
based on fitting the mass spectra of the two decay modes \hgg\ and \hzzllll.\footnote{Throughout
this paper, the symbol $\ell$ stands for electron or muon.} In these two channels the Higgs boson
produces a narrow mass peak with a typical experimental resolution of 1.6~GeV to 2~GeV over a smooth
background, from which the mass can be extracted without assumptions on the signal production and
decay yields. Interference effects are expected between the Higgs boson signal and SM background
processes.  For the \hZZllll\ channel, the impact of this interference on the mass spectrum is
negligible if the Higgs boson width is close to the SM value ~\cite{Kauer:2012hd}.  For the \hgg\
channel, such effects are larger for widths close to the SM
value~\cite{Dixon:2003yb,Dixon:2013haa,Martin:2012xc}, but still small compared to the present
experimental precision.  The interference effects on the mass spectra are neglected in this paper.

Recent measurements of the Higgs boson mass from the ATLAS and CMS Collaborations are reported in
Refs.~\cite{ATLAScouplings} and~\cite{CMSfourl}.  The ATLAS measurement was based on the same data
sample as that analyzed in this paper, corresponding to an integrated luminosity of 4.5\,\ifb\ at
$\sqrt{s}=7\TeV$ and of 20.3\,\ifb\ at $\sqrt{s}=8\TeV$ of $pp$ collisions, taken in 2011 and 2012,
respectively.  The luminosity determination for the 2012 dataset has been improved compared to
Ref.~\cite{ATLAScouplings}.

The measurement of the Higgs boson mass is updated in this work with improved analyses of the two
channels \hgg\ and \hZZllll, as described in Secs.~\ref{sec:hsg1} and~\ref{sec:hsg2}.
The \hgg\ channel profits from an improved calibration of the energy measurements of electron and
photon candidates, which results in a sizable reduction of the systematic uncertainties on their
energy scales.
In the \hzzllll\ channel both the expected statistical uncertainty and
the systematic uncertainty on the mass measurement have been reduced with respect to the previous
publication.
The improvement of the statistical uncertainty arises primarily 
from the use of a multivariate discriminant that is designed to
increase the separation of signal from background. The systematic
uncertainty reduction comes from both the improved electromagnetic 
energy calibration and a reduction in the muon momentum scale
uncertainty, which was obtained by studying large samples of
$Z \rightarrow \mu^{+} \mu^{-}$ and $J / \psi \rightarrow \mu^{+} \mu^{-}$ decays. 

More information on the general aspects of the \hgg\ and \hZZllll\ analyses is contained in
the concurrent Refs.~\cite{ATLASggFinal,ATLAS4lfinal}, where in particular the details of the signal
and background simulation can be found.  The present measurement of the Higgs boson mass relies strongly
 upon both the calibration of the energy measurement for electrons and photons described in
Ref.~\cite{ref:run1-egamma-calib}, and the understanding of the muon momentum scale and resolution
presented in Ref.~\cite{MCPpaper2014}.

The ATLAS detector ~\cite{atlas-det} is a multipurpose detector with a forward-backward symmetric
cylindrical geometry.\footnote{ATLAS uses a right-handed coordinate system with its origin at the
nominal interaction point (IP) in the center of the detector and the $z$-axis along the beam
pipe. The $x$-axis points from the IP to the center of the LHC ring, and the $y$-axis points
upward. Cylindrical coordinates $(r,\phi)$ are used in the transverse plane, $\phi$ being the
azimuthal angle around the beam pipe. The pseudorapidity is defined in terms of the polar angle
$\theta$ as $\eta=-\ln\tan(\theta/2)$.} At small radii, the inner detector (ID), immersed in a 2~T
magnetic field produced by a thin superconducting solenoid located in front of the calorimeter, 
is made up of fine-granularity pixel and microstrip detectors. These silicon-based detectors cover
the pseudorapidity range $|\eta|<2.5$. A gas-filled straw-tube transition radiation tracker (TRT)
complements the silicon tracker at larger radii and also provides electron identification based on
transition radiation. The electromagnetic (EM) calorimeter is a lead/liquid-argon sampling
calorimeter with accordion geometry.  The calorimeter is divided into a barrel section covering
$|\eta|<1.475$ and two end-cap sections covering $1.375<|\eta|<3.2$. For $|\eta|<2.5$ it is divided
into three layers in depth, which are finely segmented in $\eta$ and $\phi$. A thin presampler
layer, covering $|\eta|<1.8$, is used to correct for fluctuations in upstream energy losses.
Hadronic calorimetry in the region $|\eta|<1.7$ uses steel absorbers and scintillator tiles as the
active medium.  Liquid argon calorimetry with copper absorbers is used in the hadronic end-cap
calorimeters, which cover the region $1.5<|\eta|<3.2$.  A forward calorimeter using copper or
tungsten absorbers with liquid argon completes the calorimeter coverage up to $|\eta|=4.9$.  The
muon spectrometer (MS) measures the deflection of muon tracks with $|\eta|<2.7$, using three
stations of precision drift tubes, with cathode strip chambers in the innermost layer for
$|\eta|>2.0$. The deflection is provided by a toroidal magnetic field with an integral of approximately 3~Tm and 6~Tm 
in the central and end-cap regions of ATLAS, respectively. The muon spectrometer is also instrumented with separate
trigger chambers covering $|\eta|<2.4$.

The outline of this paper is the following. In Secs.~\ref{sec:egamma_sys} and~\ref{sec:muon_sys}, 
the improvements in the measurement of the physics objects used for the mass measurement 
(photons, electrons and muons) are described.  
In Secs.~\ref{sec:hsg1} and \ref{sec:hsg2} a brief description of the analyses used to measure the
Higgs boson mass in the \hgg\ and \hZZllll\ channels is presented, with emphasis on the improvements
with respect to the analysis published in Ref.~\cite{ATLAScouplings}.  The statistical procedures
used for the measurement of the mass and the contributions of the different systematic uncertainties
are discussed in Sec.~\ref{sec:stat_and_sys}.  The results of the combined mass measurement and
the compatibility of the individual measurements of the two channels are reported in
Sec.~\ref{sec:results}.

\section{Photon and electron reconstruction, energy scale calibration and systematic uncertainties\label{sec:egamma_sys}}
The calibration strategy for the energy measurement of electrons and photons is described in detail
in Ref.~\cite{ref:run1-egamma-calib}.  In this section, the definitions of photon and electron
objects are given, followed by a description of their energy scale calibration.  To achieve the best
energy resolution and to minimize systematic uncertainties, the calibration and stability of the
calorimeter cell energy measurement is optimized, the relative calibration of the
longitudinal layers of the calorimeter is adjusted and a determination 
of the amount of material in front of the calorimeter is performed.
The global calorimeter energy scale is then determined in situ with a
large sample of $\ztoee$ events, and verified using $\Jpsitoee$ and
$\ztollg$ events. The calibration analysis uses a total of 6.6 million $\ztoee$ decays, 
0.3 million $\Jpsitoee$ decays and 0.2 million 
radiative $Z$ boson decays. Compared to the previous
publication~\cite{ATLAScouplings}, the uncertainties in the
calibration are significantly reduced by using data-driven
measurements for the intercalibration of the calorimeter layers and
for the estimate of the material in front of the calorimeter, as well
as by improving the accuracy of the in situ calibration with $\ztoee$
events~\cite{ref:run1-egamma-calib}.

\subsection{Definition of photon and electron objects}

Photon and electron candidates are reconstructed from clusters of energy deposited in the EM
calorimeter.  Candidates without a matching track or reconstructed conversion vertex in the ID are
classified as unconverted photon candidates.  Candidates with a matching reconstructed conversion
vertex or a matching track consistent with originating from a photon conversion are classified as
converted photon candidates.  Candidates matched to a track consistent with originating from
an electron produced in the beam interaction region are kept as electron candidates.

The measurement of the electron or photon energy is based on the energy collected in calorimeter
cells in an area of size $\Delta \eta \times \Delta \phi$ of $0.075\times0.175$ for electrons and
converted photons in the barrel, $0.075\times0.125$ for unconverted photons in the barrel and
$0.125\times0.125$ for electrons and photons in the end-caps.  The choice of a different area for
electrons and unconverted photons in the barrel is driven by the deflection of charged particles in
the magnetic field and bremsstrahlung in upstream material.  A multivariate regression algorithm
to calibrate electron and photon energy measurements was developed and optimized on
simulation. Corrections are made for the energy deposited in front of the calorimeter (typically
between a few~\% and 20\% of the electron energy for 100~GeV energy electrons~\cite{atlas-det}) and outside of the cluster
(around 5\%), as well as for the variation of the energy response as a function of the impact point on the calorimeter. The inputs
to the energy calibration algorithm
are the measured energy per calorimeter layer, including the presampler, $\eta$ of the cluster and
the local position of the shower within the second-layer cell corresponding to the cluster
centroid. In addition for converted photons, the track transverse momenta and the conversion radius
are used as input to the regression algorithm to further improve the energy resolution, especially
at low energy. This calibration procedure gives a 10\% improvement in the expected mass resolution
for $\hgg$ compared to the calibration used in the previous publication.  For electron and photon
candidates, the associated tracks are fitted with a Gaussian-Sum Filter to account for
bremsstrahlung energy losses~\cite{GSFElectrons}. For \hzzllll\ candidates, the resulting momentum
measurement is combined with the energy measured in the calorimeter to improve the electron energy
measurement, especially at low energy or in the transition region between the barrel and end-cap
calorimeters, where the calorimeter and ID have similar resolution.

\subsection{Cell energy calibration and stability}

The raw signal from each calorimeter cell is converted into a deposited energy using
the electronics calibration of the EM calorimeter~\cite{ref:LAr-electronics-calibration}.
The calibration coefficients are determined periodically using dedicated electronics calibration
runs and are stable in time to better than 0.1\%.  The relative calibration of the different gains
used in the readout is investigated by studying the $\ztoee$ sample, used for the global
energy scale, as a function of the electron energy and categorizing
the events according to the electronics gain used for the energy measurement, and small corrections
(typically less than a few per mille) are applied.  The corrections applied to the few percent of
channels operated at non-nominal high voltage values are verified using data. The stability of the
calorimeter response for data, both as a function of time and of instantaneous luminosity, is monitored
using electrons from $W$ or $Z$ decays and is found to be better than 0.05\%.

\subsection{Intercalibration of the different calorimeter layers}

Accurate relative intercalibration of the different layers of the EM calorimeter is critical to
achieve good linearity of the energy response. The relative calibration of the first two layers of
the EM calorimeter, which contain most of the energy deposited by electrons and photons, is
performed using muons from $Z$ boson decays by comparing their measured energy loss in data and
simulation. The use of muons allows the determination of the intrinsic relative layer calibration,
independently of uncertainties on the material in front of the EM calorimeter. Small corrections,
around 2\% on average, for the relative calibration of the two layers are derived. The uncertainty
on the relative calibration of the first two layers of the EM calorimeter varies between 1\% and 2\%
as a function of $\eta$ and is dominated by the uncertainties on the exact amount of liquid argon
traversed by the muons and by the accuracy of the simulation of the cross-talk between calorimeter
cells.  The relative calibration of the presampler layer is derived from electrons, by comparing the 
presampler energy in data and simulation as a function of the longitudinal shower development
measured in the calorimeter.  The accuracy of this calibration, which
does not depend on knowledge of the material in front of the
presampler, is better than 5\%. 

\subsection{Determination of the material in front of the EM calorimeter}

Accurate knowledge of the material in front of the EM calorimeter is required to properly
correct for the energy lost upstream of the calorimeter, which also depends on the nature of the particle (electron,
unconverted photon, converted photon) and its energy. The total amount of material in front of
the presampler layer varies from two radiation lengths (for $|\eta|$ $<$ 0.6) to about five
radiation lengths (for $|\eta|\sim 1.7$).  The amount of material in front of the calorimeter is verified using collision
data by studying the longitudinal development of electromagnetic showers, measured using the first
two layers of the calorimeter, which are intercalibrated as described above, without any
assumption about the material in front of the calorimeter. The
uncertainties given below result from the statistical accuracy of the
data and from the uncertainties in the modeling of the longitudinal
shower profiles in the calorimeter.


The material between the presampler and the first calorimeter layer is measured using unconverted
photons with low energy deposition in the presampler. Comparison of data and simulation shows that
this material is well described in the simulation with an accuracy between 0.03 and 0.05 radiation
lengths.

The integral of the material in front of the presampler is determined using the difference between
electron and unconverted photon longitudinal shower profiles. The accuracy of this measurement is
between 0.02 and 0.10 radiation lengths, depending on $\eta$. Over most of the calorimeter acceptance,
the simulation is found to reproduce the data well, after some improvements in the description of
the material in front of the end-cap calorimeter, with the exception of a few small localized
regions where differences of up to 0.3 radiation lengths remain.  The relative calibration of
electron and photon energy measurements also depends on the radial
position of detector material in front of the presampler, which cannot
be directly probed using longitudinal shower profiles measured in the
calorimeter.  The uncertainty on the amount of material in the ID active area is
estimated from a comparison between a bottom-up inventory of the ID
components and the measured weight of different ID subdetector units~\cite{atlas-det}.
A 5\% relative uncertainty, corresponding to 0.02 to 0.10 radiation 
lengths depending on the detector region, in the amount of material in the ID active area is derived
from this comparison.  
Measurement of the rates of hadronic interactions~\cite{HadronicInteraction} and of photon conversions with collision data,
are consistent with (albeit less precise than) this a priori knowledge.
The determination of the integral of the material in front of the presampler is then
used together with knowledge of the material in the ID active area to constrain the material in
the detector services beyond the active part of the ID and in the calorimeter cryostats.

\subsection{Global calorimeter energy scale adjustment}

The global calorimeter energy scale is determined from $\ztoee$ decays by comparing the
reconstructed mass distributions in data and simulation. This is done in bins of $\eta$
of the electrons. The energy scale correction factors are typically of the order of 1--3\% and are
consistent with the uncertainties on the initial energy scale derived from test-beam data. The
uncertainty in the measurement of these factors from the $Z$ sample is less than 0.1\% on average,
and up to 0.3\% for $|\eta|\sim 1.5$ at the barrel/end-cap transition.  The uncertainty is
significantly reduced compared to Ref.~\cite{Aad:2011mk}, owing to the improved detector description
discussed above, to improved simulation, to the intercalibration corrections and to a larger $Z$ boson decay
sample.  At the same time, an effective constant term for the calorimeter energy resolution is
extracted by adjusting the width of the reconstructed $Z$ mass distribution in simulation to match
the distribution in data. This constant term is on average 0.7\% for $|\eta|<0.6$, and between 0.7\%
and 1.5\% in the remainder of the calorimeter acceptance, except in the transition region between
barrel and end-cap calorimeters where it is 3.5\% and at the end of the end-cap acceptance
($|\eta|>$2.3) where it is 2.5\%.  This constant term is used to adjust the energy resolution in
simulated samples.  The extraction of the energy scale and of the effective constant term is done
separately for the 7~TeV and 8~TeV data. The effective constant term is about 0.2--0.3\% larger in
the 8~TeV data.

\subsection{Systematic uncertainties on the energy scale and cross-checks}

The calorimeter energy scale adjustment with $Z$ events determines the scale for electrons with
transverse energy (\eT) close to that of $\ztoee$ events ($\et\sim
40$~GeV on average). 
Any systematic uncertainty has thus minimal impact for 40~GeV \et\ electrons but can lead to residual non-linearities 
and differences between the electron, unconverted photon and converted photon energy scales.

In addition to the uncertainty on the overall energy scale adjustment, the uncertainties affecting
the energy measurement of electrons and photons can be classified as
follows. The impact of these systematic uncertainties on the photon
energy scale is detailed for photons from Higgs boson decays, as the
impact of energy scale systematic uncertainties is larger for this decay channel.

\begin{itemize}

\item Uncertainty on the non-linearity of the energy measurement at the cell level: this arises
  mostly from the relative calibration of the different gains used in the calorimeter
  readout.  The uncertainty on the non-linearity of the cell energy 
  calibration contributes an uncertainty of about 0.1\% to the energy scale of photons from
  Higgs boson decays (up to 1\% for $1.5<|\eta|<1.7$).

\item Uncertainty on the relative calibration of the different calorimeter layers: these
  contribute an uncertainty of about 0.10\% to 0.15\% to the energy
  scale of photons from Higgs boson
  decays.

\item Uncertainty on the amount of material in front of the calorimeter: these
  contribute between 0.1\% and 0.3\% as a function of $\eta$ for unconverted photons from Higgs
  boson decays. This uncertainty is typically two times smaller for
  converted photons that have an
  energy loss before the calorimeter closer to that of the $Z$ decay electrons used in the energy
  scale adjustment.

\item Uncertainty in the reconstruction of photon conversions: unconverted and converted photons are
  calibrated differently to take into account the difference in the energy loss before the
  calorimeter. 
  Converted photons misidentified as unconverted photons, or vice-versa,  are typically reconstructed with an energy shifted by 2\%. The
  uncertainty in the modeling of the efficiency to properly classify converted or
  unconverted photons is a few percent. This translates into an uncertainty on the photon energy
  scale of 0.02--0.04\% for both the converted and unconverted photons.

\item Uncertainty in the modeling of the lateral shower shape: differences between data and simulation
  for the lateral development of electromagnetic showers contribute to the uncertainty on the energy scale
  if they depend on energy or particle type. These differences are compared for photons and electrons using a sample of
  radiative $Z$ decays. They are found to be consistent. The resulting uncertainty on the photon energy
   scale is 0.05--0.3\% 
  depending on $\eta$ and whether or not the
  photon converted.
\end{itemize}

At an \eT\ of about 60~GeV, the total uncertainty on the photon energy scale is between
0.2\% and 0.3\% for $|\eta|<1.37$ or $|\eta|>1.82$;  for $1.52<|\eta|<1.82$, the
uncertainty is 0.9\% and 0.4\% for unconverted and converted photons, respectively.
The energy dependence of the photon energy scale uncertainty is weak.
The uncertainty on the electron energy scale at an \et\ of 40~GeV is on average 0.03\% for
$|\eta|<1.37$, 0.2\% for $1.37<|\eta|<1.82$ and 0.05\% for $|\eta|>1.82$.
At an \et\ of about 10~GeV, the electron energy scale uncertainty ranges from 0.4\% to 1\%
for $|\eta|<1.37$, is about 2\% for $1.37<|\eta|<1.82$, and again 0.4\% for $|\eta|>1.82$.
The largest uncertainty for electrons is in the
barrel/end-cap transition region, which is not used for photons. These
uncertainties are modeled using 29 independent sources to
account for their $\eta$-dependence, and are almost fully correlated
between the 7~TeV and 8~TeV samples.

\begin{figure*}
          \centering
          \mbox{  
            \begin{tabular}{cc}
             \subfigure[\label{fig:scaleelec_1}]{\includegraphics[width=0.42\linewidth]{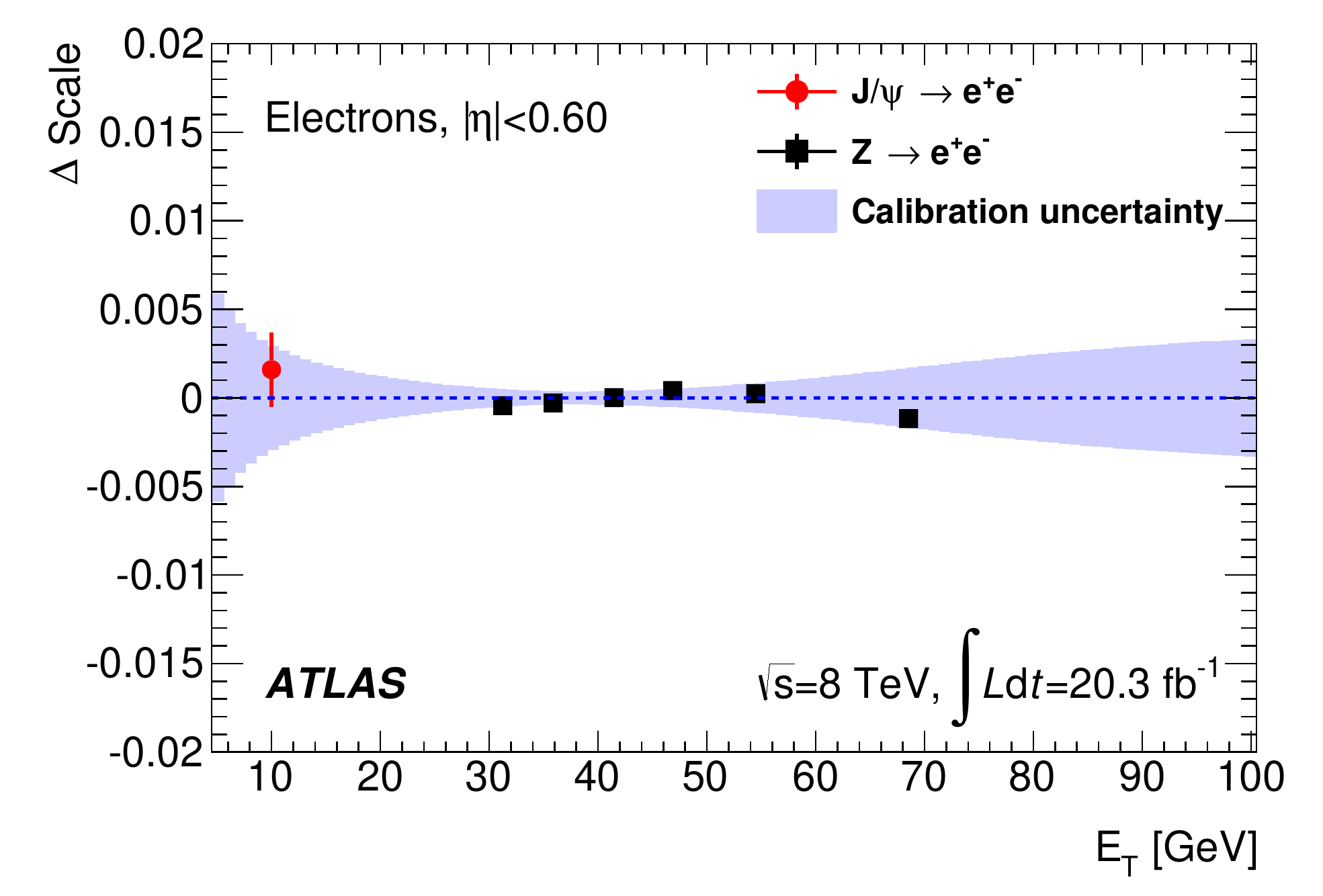}} &
             \subfigure[\label{fig:scaleelec_2}]{\includegraphics[width=0.42\linewidth]{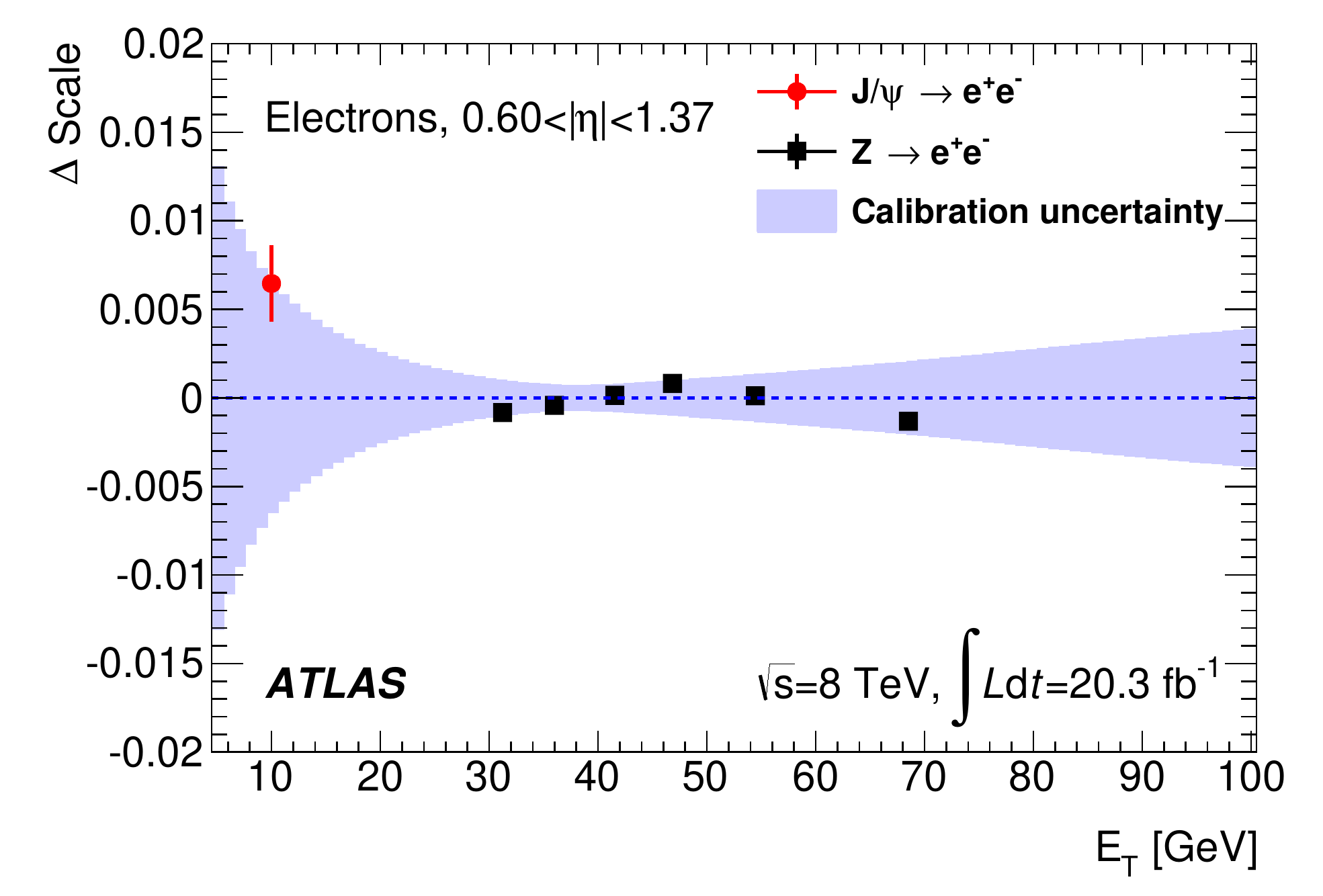}} \\
             \subfigure[\label{fig:scaleelec_3}]{\includegraphics[width=0.42\linewidth]{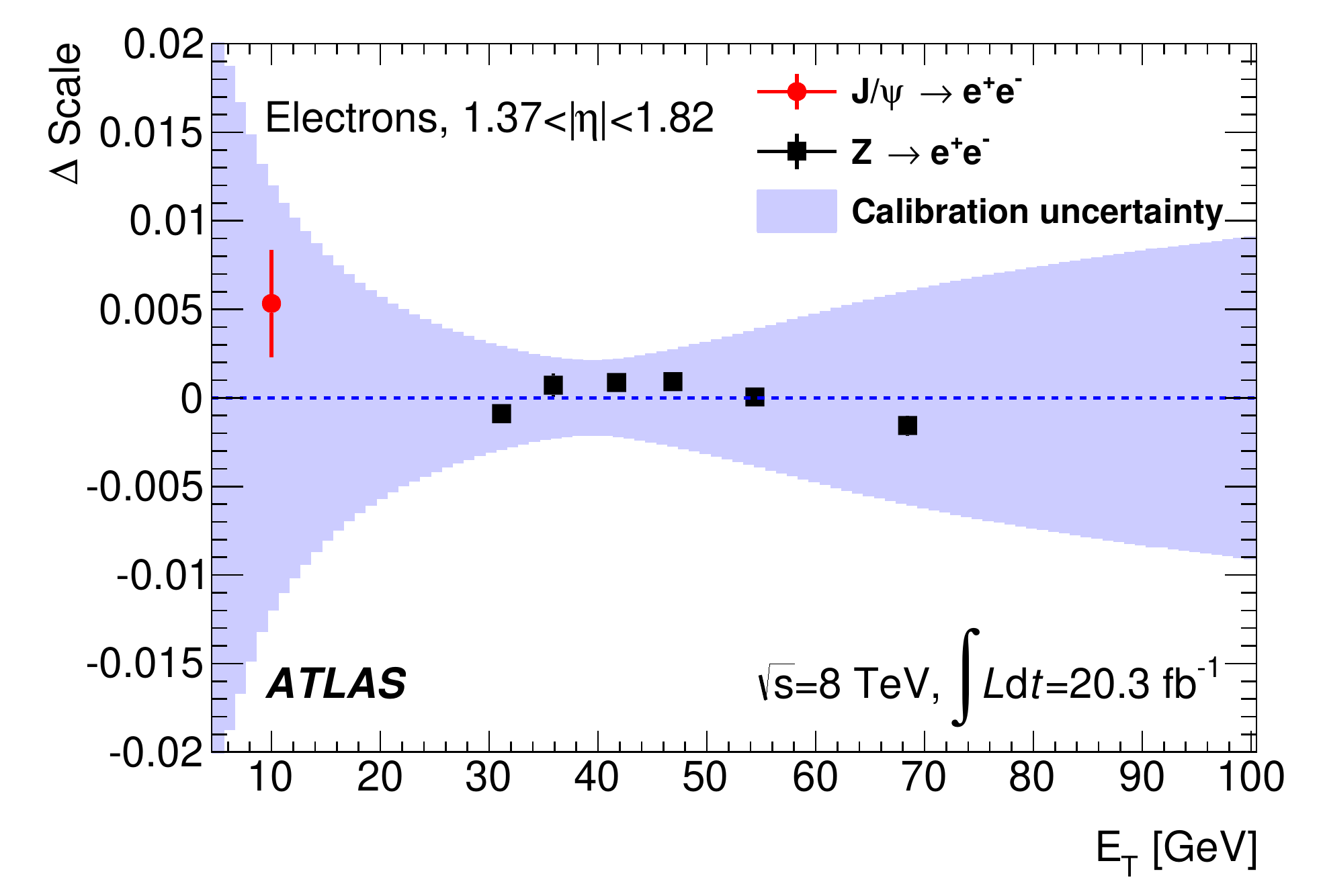}} &
             \subfigure[\label{fig:scaleelec_4}]{\includegraphics[width=0.42\linewidth]{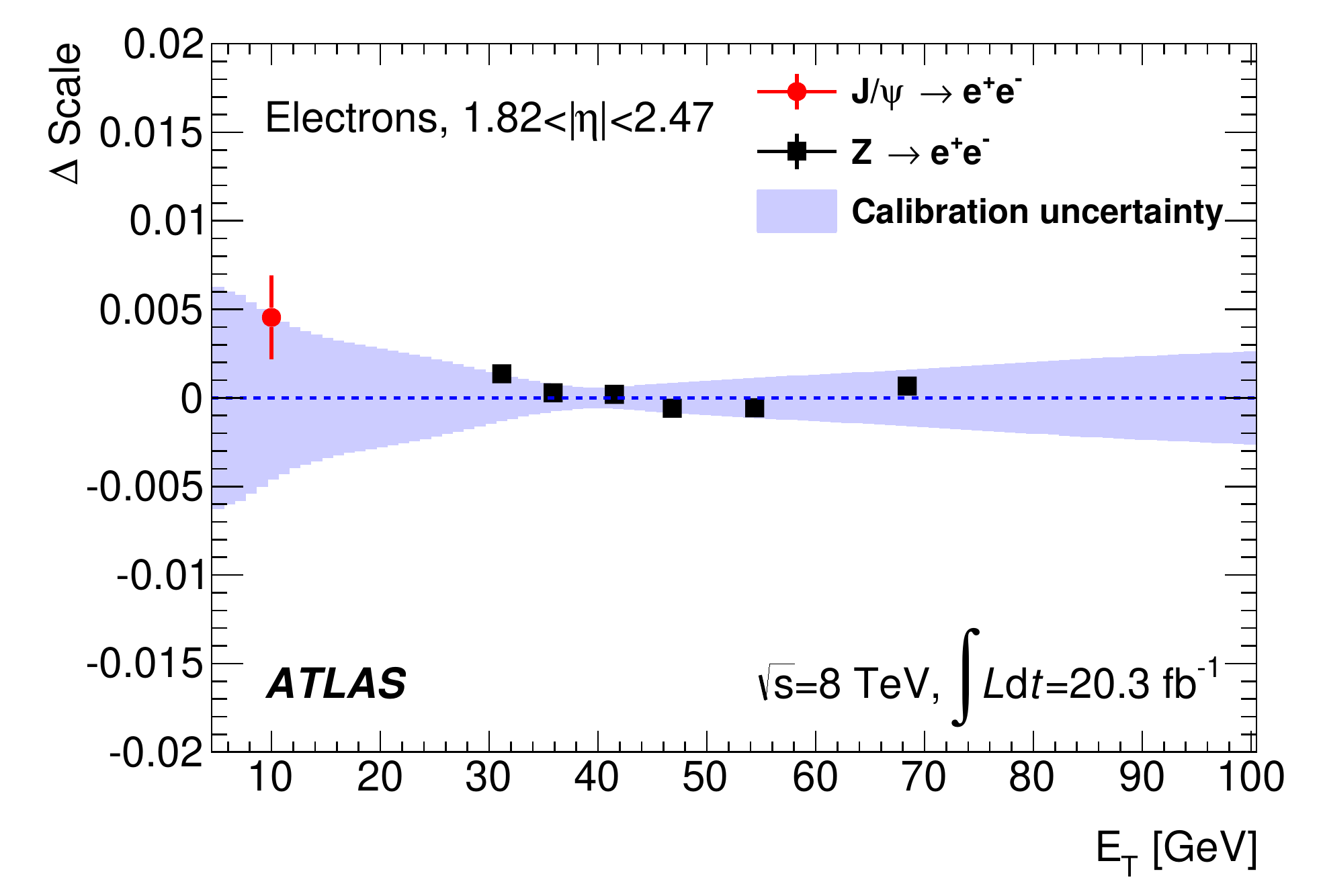}} 
            \end{tabular}
               }       
           \caption{Relative scale difference, $\Delta$~Scale, between the measured electron energy scale and the
             nominal energy scale, as a function of \et\ using
             $\Jpsitoee$ and $\ztoee$ events (points with error bars), for
             four different $\eta$ regions:
             \subref{fig:scaleelec_1} $|\eta|<0.6$, \subref{fig:scaleelec_2} $0.6<|\eta|<1.37$, 
             \subref{fig:scaleelec_3}  $1.37<|\eta|<1.82$ and
             \subref{fig:scaleelec_4}  $1.82<|\eta|<2.37$.
             The uncertainty on
             nominal energy scale for electrons is shown as the shaded area. The error bars include the systematic
             uncertainties specific to the  $\Jpsitoee$ measurement.}
           \label{fig:egamma_scales_elec}
\end{figure*}

\begin{figure*}
          \centering
          \mbox{  
            \begin{tabular}{cc}
             \subfigure[\label{fig:scalephot_1}]{\includegraphics[width=0.42\linewidth]{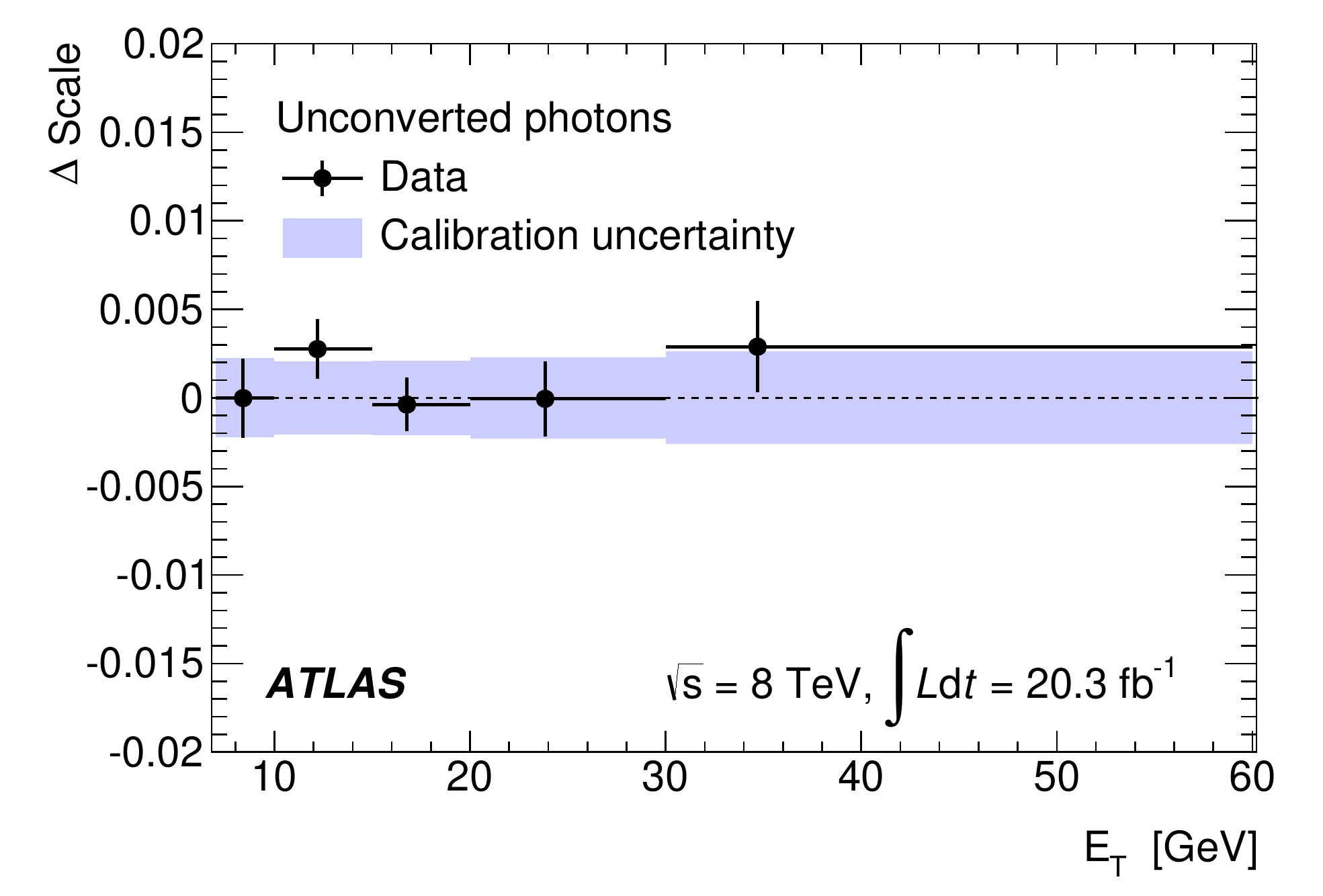}} &
             \subfigure[\label{fig:scalephot_2}]{\includegraphics[width=0.42\linewidth]{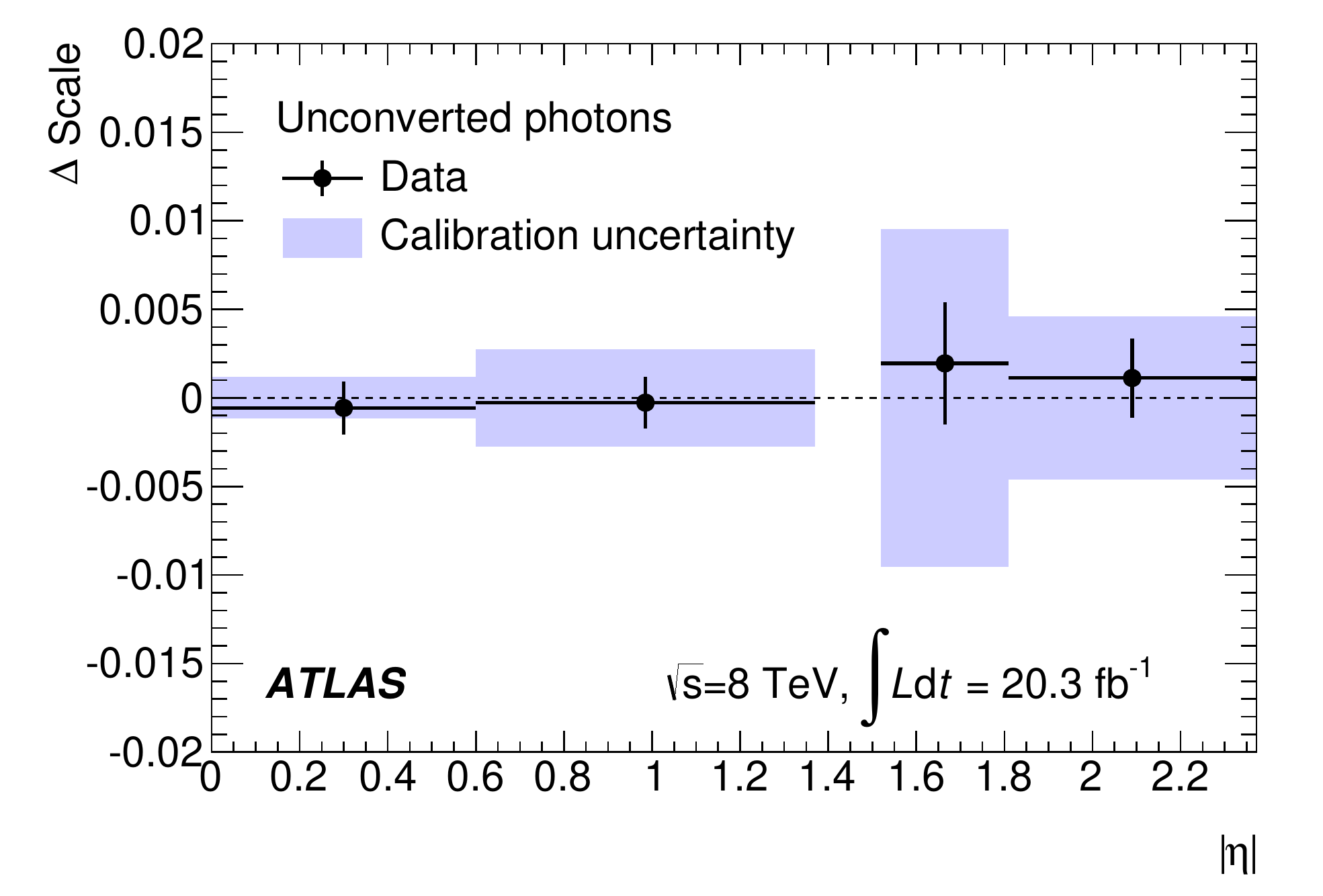}} \\
             \subfigure[\label{fig:scalephot_3}]{\includegraphics[width=0.42\linewidth]{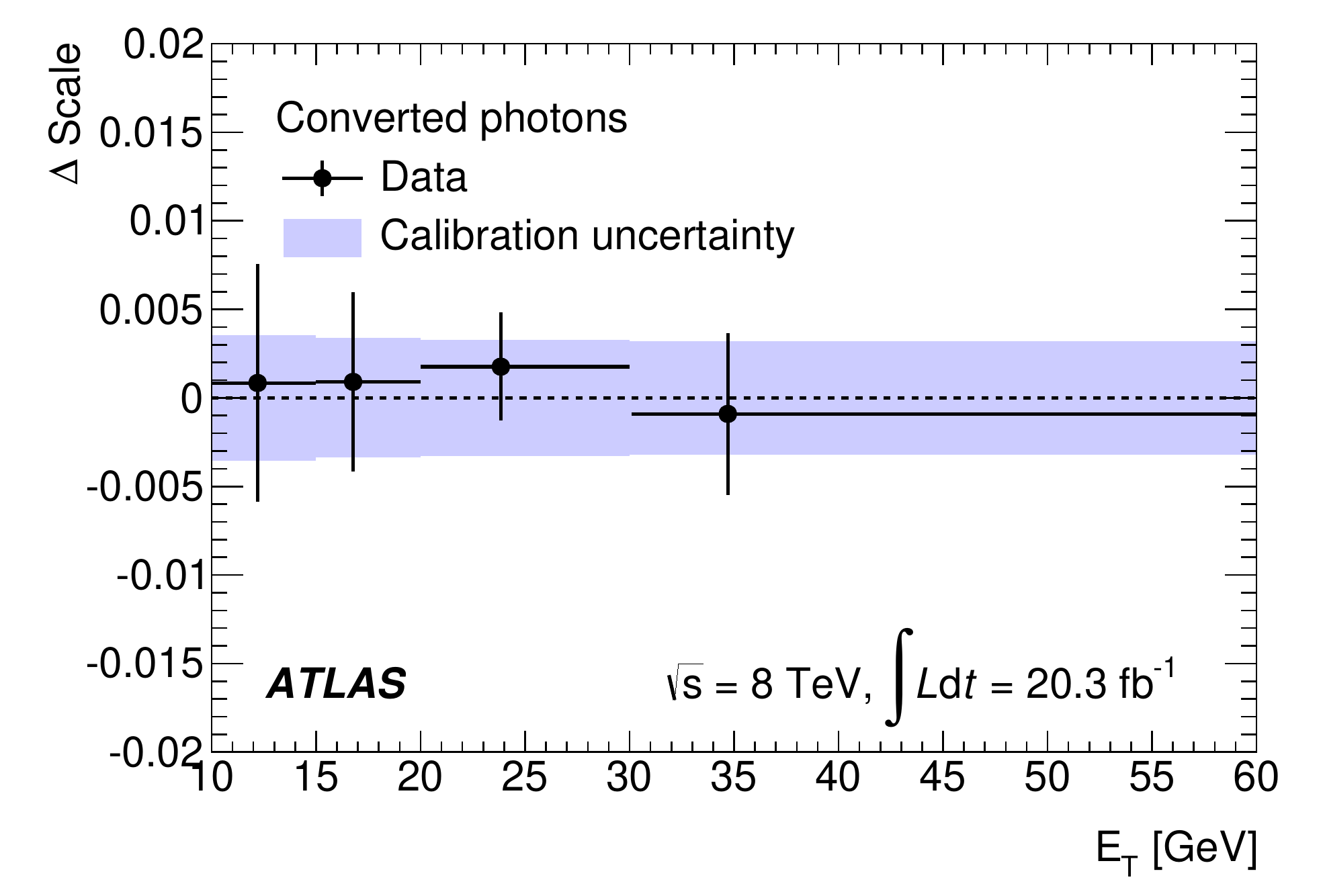}} &
             \subfigure[\label{fig:scalephot_4}]{\includegraphics[width=0.42\linewidth]{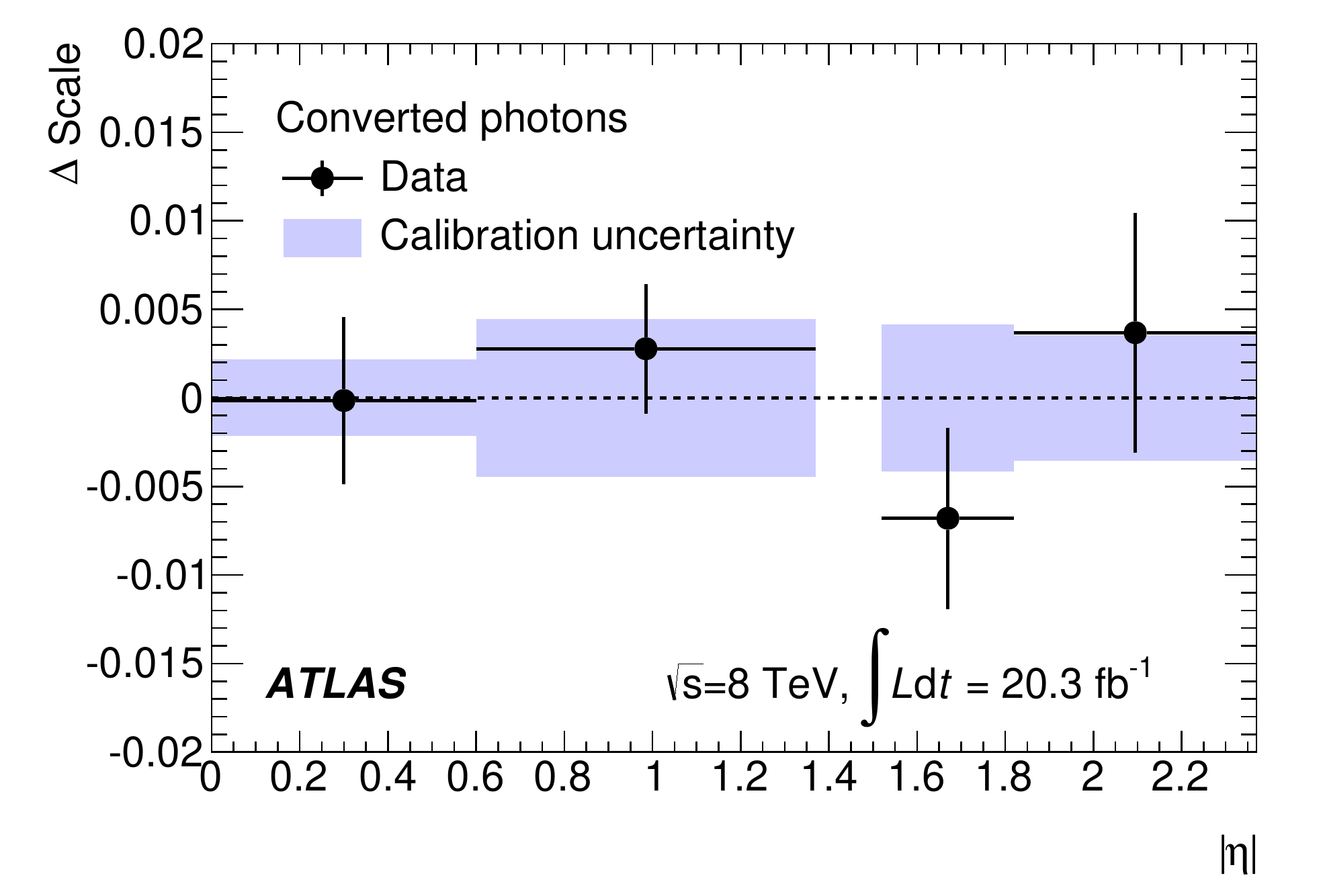}} \\
            \end{tabular}
               }           
           \caption{Relative scale difference, $\Delta$~Scale, between the measured photon energy scale using $Z\rightarrow
             \ell\ell\gamma$ events and the nominal energy scale: \subref{fig:scalephot_1} as a function of \et\
             for unconverted photons, \subref{fig:scalephot_2} as a function of $\eta$ for unconverted photons,
             \subref{fig:scalephot_3} as a function of \et\ for converted photons and
             \subref{fig:scalephot_4} as a function of $\eta$ for converted photons.
             Photons reconstructed in the
             barrel/end-cap transition region are not considered. The
             $Z\rightarrow\ell\ell\gamma$ measurements are the points
             with error bars. The uncertainty on the 
             nominal energy scale for photons is shown as the shaded area. The error bars include the systematic uncertainties
             specific to the $Z\rightarrow\ell\ell\gamma$ measurement.}
           \label{fig:egamma_scales_phot}
\end{figure*}

An independent verification of the energy scale is performed using samples of $\Jpsitoee$
and $\ztollg$ decays. The latter sample allows, for instance, a direct measurement of the photon energy
scale in the low transverse energy range (typically between 7~GeV and
35~GeV). The results are in good agreement with the energy scale determined from the $\ztoee$ sample, taking into account the
systematic uncertainties discussed above. With the $\ztollg$ sample, the energy scale of photons 
with transverse energy around 30~GeV is probed with an accuracy of about 0.3\%.
Figures~\ref{fig:egamma_scales_elec} and ~\ref{fig:egamma_scales_phot}
summarize the verifications of the electron and photon energy scales from these
samples using the 8~TeV dataset, after the full calibration procedure
is applied.  In addition to the $\Jpsitoee$ and $\ztollg$ samples, the
non-linearity in the electron energy scale is also probed by dividing
the $\ztoee$ sample into bins of electron \et. These figures also show the
total systematic uncertainty on the electron and photon energy scales
as a function of \et\ and $\eta$.  The same verifications are performed using the
7~TeV dataset with results consistent within uncertainties.

\subsection{Uncertainties on the calorimeter energy resolution}

Systematic uncertainties on the calorimeter energy resolution arise
from uncertainties in the modeling of the sampling term and on the
measurement of the constant term in $Z$ boson decays, from uncertainties
related to the amount of material in front of the calorimeter,
which affect electrons and photons differently, and from 
uncertainty in the modeling of the small contribution to the resolution from fluctuations in the
pile-up from other proton--proton interactions in the same or neighboring bunch crossings. The
uncertainty on the calorimeter energy resolution is typically
$\sim$10\% for photons from Higgs boson decays, and varies from 10\%
to 5\% for electrons in the \et\ range from 10~GeV to 45~GeV.

\section{Muon reconstruction, momentum scale and resolution systematic uncertainties \label{sec:muon_sys}}
The muon momentum is measured independently by the ID and the MS detector systems.  Four types
of muon candidates are reconstructed, depending on the available information from the ID, the MS,
and the calorimeters.  Most muon candidates are identified by matching a reconstructed
ID track with either a complete or a partial (local segment) track reconstructed in the
MS~\cite{MCPpaper2014, MCPpaper2010}.
If a complete MS track is
present, the two independent momentum measurements are combined (CB muons),
otherwise the momentum is measured using the ID and the partial MS track serves as identification
(segment-tagged muons). The muon reconstruction and identification coverage is extended by
using tracks reconstructed in the forward region ($2.5<|\eta|<2.7$) of the MS, which is outside
the ID coverage (standalone muons). The parameters of the muon track reconstructed in the MS 
are expressed at the interaction point by extrapolating the track back to the point of closest 
approach to the beam line, taking into account the energy loss of the muon in the calorimeters.  
In the center of the barrel region ($|\eta|< 0.1$), which
lacks MS geometrical coverage, ID tracks with transverse momentum $\pt>15$~GeV are identified as muons if their
calorimetric energy deposits are consistent with a minimum ionizing particle
(calorimeter-tagged muons). 
The combination of the track measurements provided by the ID and MS ensures excellent 
momentum resolution across three orders of magnitude, from a few GeV
up to a few TeV.  

The muon reconstruction in simulation is corrected to match the momentum scale and
resolution measured from collision data as described in detail in Ref.~\cite{MCPpaper2014}.
About 6 million $\Jpsimm$ events\footnote{Only a subset of the   
$J/\psi$ events are used to derive the muon momentum corrections, in order to balance the weight of 
the $\Jpsimm$ and $\Zmm$ events.} and about 9 million $\Zmm$ events were used to extract the corrections 
to be applied to the simulated data.
They consist of scale corrections for the ID and MS, a $p_{\rm T}$-independent momentum correction for the MS  
and a $p_{\rm T}$-dependent smearing correction to be applied to
reproduce the resolution observed in data.  The corrections for the ID
and MS momentum measurements were derived separately.  
For the momentum of CB muons, the individual corrections from the ID and MS momentum are 
combined according to their relative weight in the measurement of the combined muon.

To extract the ID corrections, template fits to the $\Jpsimm$ and $\Zmm$ invariant mass
distributions are performed in bins of $\eta$ and $p_{\rm T}$. The MS corrections are extracted by
fitting the $\Jpsimm$ and $\Zmm$ invariant mass distributions and the
difference between the momentum measured in the ID and MS. The MS
corrections are derived in bins of $p_{\rm T}$ and $\eta$, and follow
the sector 
granularity of the MS in the azimuthal coordinate $\phi$.  The systematic uncertainties on the
corrections are estimated by varying several ingredients of the fit procedure: the
parameterization and the normalization of the backgrounds, the fit ranges, and the parameterization of
the resonances and their kinematic distributions.  The systematic uncertainties on the resolution
are varied independently for the ID and MS, whereas the ID and MS systematic scale
uncertainties are treated as fully correlated, hence maximizing the impact of the scale
variation on the CB muons.

The major improvement with respect to the previous publication is the use of $\Jpsimm$ events in addition
to the $\Zmm$ sample in the simulation correction procedure. This allows a significant reduction of the
momentum scale uncertainty in the low momentum range that is relevant for the \hZZllll\ mass measurement.  In
previous studies, the $\Jpsimm$ sample was used only for the evaluation of the systematic
uncertainties.

The ID momentum scale corrections are below $0.1\%$. The systematic uncertainties on the ID scale
increase with $|\eta|$, starting from $0.02\%$ at $\eta=0$ and rising to about $0.2\%$ for
$|\eta| >2$.  The MS scale corrections vary from $-0.4\%$ to $+0.3\%$ depending on the $\eta$ and 
$\phi$ regions. The $p_{\rm T}$-independent momentum correction to the MS measurement takes into account 
the difference between the muon energy loss in the calorimeters
in data and simulation, is of the order of a few tens of MeV and has a
negligible impact on the Higgs boson mass measurement.  Typical systematic uncertainties on the MS momentum scale
range from less than $0.1\%$ to about $0.2\%$.  The systematic uncertainties on the CB momentum scale
are $0.04\%$ in the barrel region and increase to about $0.2\%$ for $|\eta|>2$.

These results were checked by separately fitting the dimuon invariant mass distribution to extract
the peak position and the width of the $J/\psi$, $Z$ and $\Upsilon$ resonances in data and in the
simulation, with and without corrections.  For this study 17 million $J/\psi$ events were
used. The $\Upsilon$ sample, about 5 million events, was not used in the simulation correction
procedure and therefore provides an independent validation performed in bins of $p_{\rm T}$, $\eta$
and $\phi$. Figure~\ref{fig:scaleeta} shows the ratio of the reconstructed dimuon invariant mass for
data to the corrected mass in simulation for $J/\psi$, $\Upsilon$ and $Z$ events as a function of
$\eta$ of the higher-$p_{\rm T}$ muon.  Figure~\ref{fig:scalept} shows the same ratio as a function
of the average transverse momentum, $\langle \pt \rangle$, of the two muons.  The error bars on data
points show the combined statistical and systematic uncertainty. The systematic uncertainty is
extracted by varying the fitted dimuon mass range and in the case of $J/\psi$ by taking into account
the uncertainty on the background.  These studies demonstrate the validity of the corrections and of
the associated systematic uncertainties in the range $6 < p_{\rm T} \lesssim 100$ \gev.

\begin{figure}[htbp!]
  \centering
  \subfigure[\label{fig:scaleeta}]{\includegraphics[width=\largeSinglePlotSize]{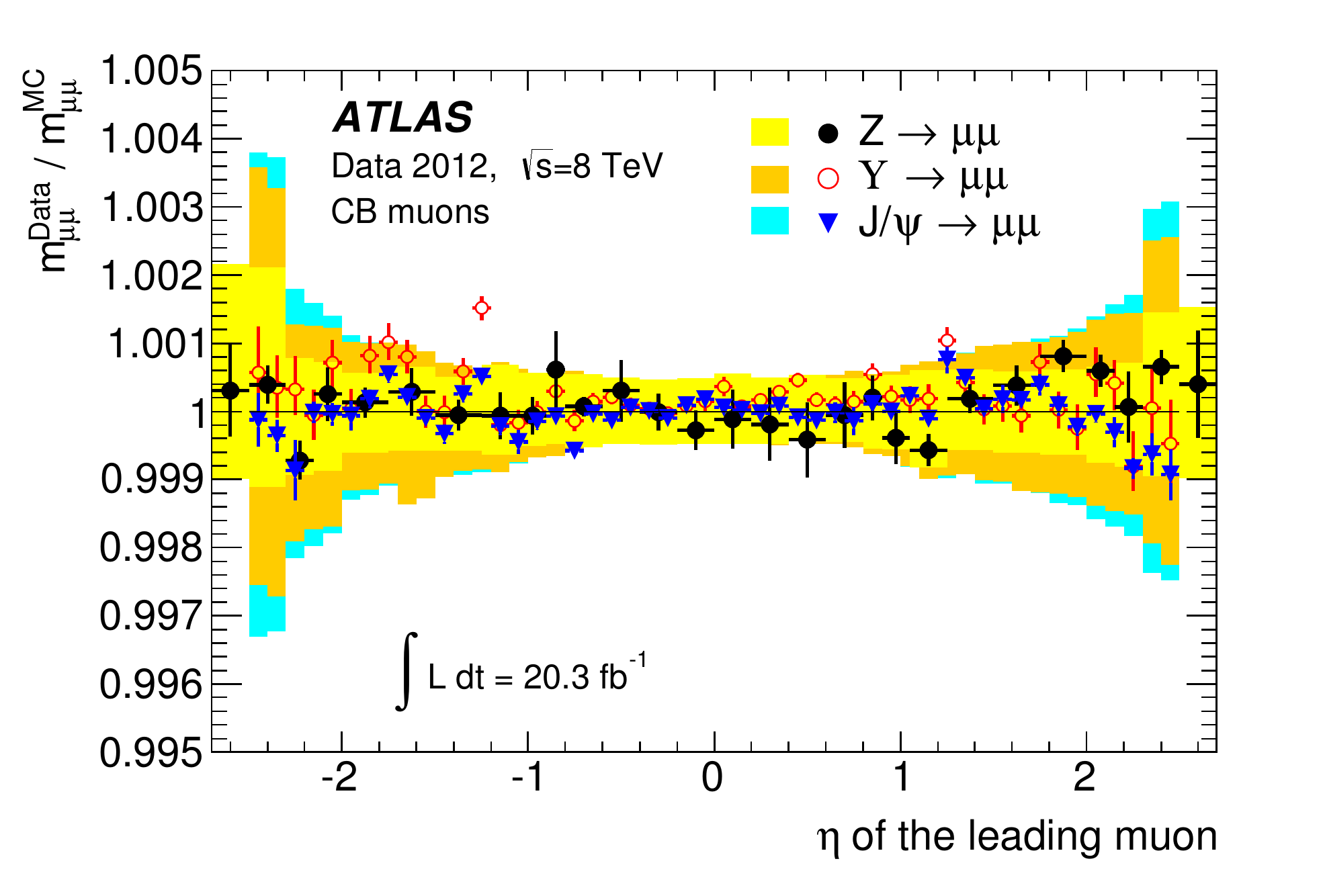}}
  \subfigure[\label{fig:scalept}]{\includegraphics[width=\largeSinglePlotSize]{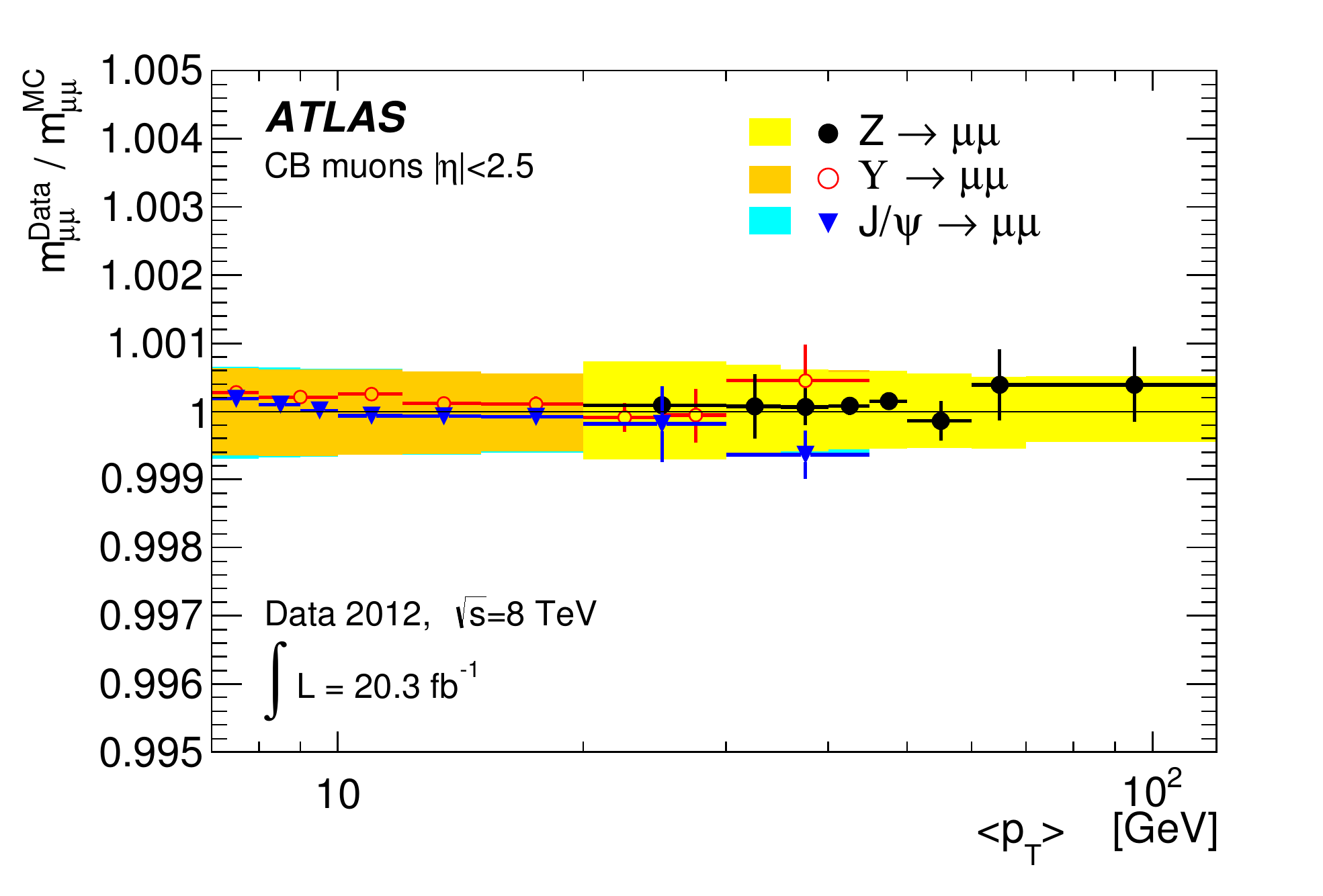}}
  \caption{ Ratio of the reconstructed dimuon invariant mass for data to the corrected mass in simulation for
    $J/\psi$, $\Upsilon$ and $Z$ events:~\subref{fig:scaleeta} as a function of $\eta$ of the
    higher-$p_{\rm T}$ muon, and~\subref{fig:scalept} as a function of $\langle \pt \rangle$ of the two
    muons, as defined in the text. The shaded areas show the systematic uncertainty on the simulation corrections
    for each of the three samples. The error bars on the points show the combined statistical and systematic uncertainty 
    as explained in the text.\label{fig:scaleEtaPt}}
 \end{figure}

\section{Mass and width measurement in the \hgg\ channel \label{sec:hsg1}}
The $H\rightarrow\gamma\gamma$ channel provides good sensitivity to the Higgs boson mass,
due to the excellent mass resolution in the diphoton final state, allowing the
observation of a narrow mass peak over a smooth background which can be determined directly from
data. The EM calorimeter provides a measurement of the photon energy and direction (photon pointing), utilizing its longitudinal segmentation.
The typical mass resolution is 1.7~GeV for a 125~GeV Higgs boson mass.
The main background is continuum $\gamma\gamma$ production with smaller contributions, of about 20\%, from the $\gamma$+jet and
dijet processes.
A more complete description of the selection criteria and background modeling is
reported in Ref.~\cite{ATLASggFinal}.

\subsection{Event Selection}
\label{sec:hsg1_evsel}

Events are selected using a diphoton trigger. For the 7~TeV data, an
\et\ threshold of 20~GeV is applied to both photons at the trigger level. 
For the 8~TeV data, the \et\ threshold at the trigger level is
35~GeV for the photon with highest \et\ and 25~GeV for the photon with next-highest \et. Loose
photon identification cuts are applied at the trigger level, which is more than 99\% efficient for
events fulfilling the final analysis selection.

Only photon candidates with $|\eta| < 2.37$ are considered, removing
the transition region $1.37<|\eta|<1.56$ between the barrel and end-cap calorimeters.  The calorimeter
granularity in the transition region is reduced, and the presence of significant additional inactive
material affects the identification capabilities and energy resolution.

Two photons are required to fulfill tight identification criteria that are based primarily on
shower shapes in the EM calorimeter~\cite{ATLAS-CONF-2012-123}. For the 7~TeV data, a neural network
discriminant is built from shower shape variables to suppress the contamination from jets
misidentified as photon candidates.
For the 8~TeV data, a set of cuts optimized for the pile-up conditions of the 2012 data taking are applied.
The efficiency of the photon identification selection ranges between 85\% and 95\% as a
function of the photon \et.

To further reject background from jets misidentified as photons, the photon candidates are required
to be isolated using both the calorimeter isolation and track isolation requirements.  The
calorimeter isolation is defined as the sum of the \et\ of clusters of energy deposited in a cone
of size $\Delta R$=$\sqrt{(\Delta\eta)^2 + (\Delta\phi)^2}=0.4$ around
the photon candidate, excluding an area of size
$\Delta \eta \times \Delta \phi = 0.125\times0.175$ centered on the
photon cluster; the expected photon energy deposit outside the
excluded area is subtracted. The pile-up and underlying event contribution to the
calorimeter isolation is subtracted
event-by-event~\cite{ref:hsg1-ambientCorrection}.  The calorimeter
isolation is required to be smaller than 5.5~GeV for the 7~TeV data,
and smaller than 6~GeV for the 8~TeV data. The track isolation is
defined as the scalar sum of the transverse momenta of the tracks in a 
cone of $\Delta R$=0.2 around the photon candidate. The tracks are required to have $\pt > 0.4~
(1.0)$~GeV, for the 7~(8)~TeV data and to be consistent with originating from the diphoton primary vertex, defined
below.  In the case of converted photons, the tracks associated with the photon conversion are
excluded from the track isolation.  The track isolation is required to be smaller than 2.2~GeV for
the 7~TeV data and smaller than 2.6~GeV for the 8~TeV data.  The efficiency of the isolation
requirement is about 95\% per photon for both 7~TeV and 8~TeV data.

Identifying which reconstructed primary vertex corresponds to the $pp$ collision that produced the
diphoton candidate is important for the mass reconstruction. The correct identification 
of the tracks coming from the $pp$ collision producing the diphoton candidate is also necessary to avoid
pile-up contributions to the track isolation.  To keep the contribution of the opening angle
resolution to the mass resolution significantly smaller than the energy resolution contribution, a
position resolution for the primary vertex of about 15 mm in the $z$ direction is sufficient.
Better resolution is needed to correctly match tracks to the $pp$ collision vertex of the
diphoton candidate.  The directions of the photon candidates can be measured using the longitudinal
and transverse segmentation of the EM calorimeter.  An estimate of the diphoton primary vertex $z$
position is obtained by combining the average beam-spot position with this photon pointing, which can be enhanced by
using the tracks from photon conversions with conversion radii before
or in the silicon detectors. This estimate gives a resolution of about
15~mm in the $z$ direction. 
In order to select the best reconstructed primary vertex, three
additional variables are defined for each reconstructed primary vertex: $\Sigma \pt$ of the track
transverse momenta, $\Sigma \pt^2$ and the azimuthal angle between 
the combined photon system and the combined system of the tracks in the transverse plane.
A neural network discriminant is constructed using both the diphoton primary vertex $z$ position estimated
by the photon pointing with its uncertainty and
this additional track information to select the best primary vertex
candidate for the diphoton event.  This algorithm selects a primary vertex within $\pm$15~mm
in $z$ of the true production vertex with an efficiency of 93\% for the average
pile-up conditions in the 8~TeV dataset. The contribution of the opening angle resolution
to the mass resolution is thus negligible.

The diphoton invariant mass $m_{\gamma\gamma}$ is computed using the measured photon energies and
their opening angle estimated from the selected primary vertex and the photon impact points in the
calorimeter.
The transverse energy is required to be $\et > 0.35\times m_{\gamma\gamma}$ for the photon with highest \et\
and $\et > 0.25\times m_{\gamma\gamma}$ for the photon with second highest $\et$. This selection leads to a
smoother background distribution in each of the event categories compared to using
fixed cuts on \et.  The combined signal reconstruction and selection efficiency for the
Higgs boson signal at an assumed mass of 125~GeV is around 40\%.
In total 94627 (17225) events are selected in the 8~TeV (7~TeV) dataset with
$105<m_{\gamma\gamma}<160$~GeV.

\subsection{Event Categorization}
\label{sec:hsg1_evcat}

To improve the accuracy of the mass measurement, the selected events are separated into ten mutually
exclusive categories that have different signal-to-background ratios, different diphoton
invariant mass resolutions and different systematic uncertainties. To keep the analysis simple, the
categorization is based only on the two photon candidates.  The categorization,
which is different from the one used in Ref.~\cite{ATLASggFinal}, is optimized to
minimize the expected uncertainty on the mass measurement, assuming a Higgs boson signal produced with the
predicted SM yield, while also accounting for systematic uncertainties.  
Events are first separated into two groups, one where both photons are
unconverted and the other where at least one photon is converted. The energy resolution for
unconverted photons is better than the one for converted photons, and energy
scale systematic uncertainties are different for converted and unconverted photons.
The events are then classified
according to the $\eta$ of the two photons: the {\it central} category corresponds to events where
both photons are within $|\eta|<0.75$, the {\it transition} category corresponds to events with at
least one photon with $1.3<|\eta|<1.75$, and the {\it rest} category corresponds to all other diphoton
events. The {\it central} category has the best mass resolution and signal-to-background
ratio, as well as smallest energy scale uncertainties. 
The {\it transition} category suffers from worse energy resolution, due to the larger amount
of material in front of the calorimeter, and also from larger systematic uncertainties.  Finally,
the {\it central} and {\it rest} categories are each split into a low \ptt\ ($<70$~GeV) and a high
\ptt\ ($>70$~GeV) category, where \ptt\ is the component of the diphoton transverse momentum
orthogonal to the diphoton thrust axis in the transverse plane.\footnote{$p_{\rm{Tt}} = |(\vec{p}_\mathrm{T}^{\gamma_1} + \vec{p}_\mathrm{T}^{\gamma_2}) \times
  \hat{\bf t}|$, where $\hat{\bf t} = \frac{ {\vec{p}_\mathrm{T}^{\gamma_1}}
    -{\vec{p}_\mathrm{T}^{\gamma_2}} } { |{\vec{p}_\mathrm{T}^{\gamma_1}} -
    {\vec{p}_\mathrm{T}^{\gamma_2}}|}$ is the thrust axis in the transverse plane, and
  ${\vec{p}_\mathrm{T}^{\gamma_1}}$, ${\vec{p}_\mathrm{T}^{\gamma_2}}$ are the transverse momenta of
  the two photons.} The high \ptt\ categories have better
  signal-to-background ratios and mass
resolution, but have smaller yield.  This categorization provides a 20\% reduction of
the expected statistical uncertainty compared to an inclusive measurement.

\subsection{Signal modeling}
\label{sec:hsg1_sigmod}

The signal mass spectrum is modeled by the sum of a Crystal Ball function for the bulk of the
events, which have a narrow Gaussian spectrum in the peak and tails toward lower reconstructed mass,
and a wide Gaussian distribution to model the far outliers in the mass resolution.
The Crystal Ball function is defined as:

\begin{small}
\begin{center}
\begin{displaymath}
 N \cdot \left\{ 
    \begin{array}{ll}
      e^{-t^{2}/2} & \mbox{if $t>-\alpha_{\mathrm{CB}}$}  \\ 
      (\frac{n_{\mathrm{CB}}}{\alpha_{\mathrm{CB}}})^{n_{\mathrm{CB}}}  e^{-\alpha_{\mathrm{CB}}^2/2}  (\frac{n_{\mathrm{CB}}}{\alpha_{\mathrm{CB}}}-\alpha_{\mathrm{CB}}-t)^{-n_{\mathrm{CB}}}  & \mbox{otherwise} 
    \end{array} 
  \right.  
\end{displaymath}
\end{center}
\end{small}
where $t = (m_{\gamma\gamma}-\mu_{\mathrm{CB}})/\sigma_{\mathrm{CB}}$, $N$ is a normalization parameter, $\mu_{\mathrm{CB}}$ is the peak of the narrow Gaussian distribution, 
$\sigma_{\mathrm{CB}}$ represents
the Gaussian resolution for the core component, and $n_{\mathrm{CB}}$ and $\alpha_{\mathrm{CB}}$ parameterize the non-Gaussian tail.

The $\sigma_{\mathrm{CB}}$ parameter varies from 1.2~GeV to 2.1~GeV depending on the category of the event.  The
overall resolution can be quantified either through its full width at half maximum (FWHM), which varies from 2.8~GeV to 5.3~GeV or
using $\sigma_\text{eff}$, defined as half of the smallest range containing 68\% of the signal events, which
varies from 1.2~GeV to 2.4~GeV.

The parameters of the Crystal Ball and Gaussian functions, and their dependence on the Higgs boson mass, 
are fixed by fits to simulation samples at discrete mass values to obtain a 
smooth signal model depending only on the assumed Higgs boson mass and yield. 
The accuracy of this procedure is checked by fitting
the Higgs boson mass in simulated samples with this signal model and is found to be better than
0.01\% of the Higgs boson mass.

\subsection{Background modeling and estimation}
\label{sec:hsg1_bkgmod}

The background is obtained directly from a fit to the diphoton mass distribution in the data over
the range 105--160~GeV after final selection. The procedure used to select the analytical
form of the function describing the background shape is explained in more detail in
Ref.~\cite{ATLASggFinal}. Different analytical forms are evaluated using a large simulated
background sample composed of diphoton events, photon+jet events (with one jet misidentified as photon)
and dijet events (with both jets misidentified as photons). Signal-plus-background fits are performed
on this background-only sample, thus the fitted signal yield should be zero if the functional form
used describes the background shape well.  The functional form retained to describe the background
is required to have a spurious fitted signal less than 20\% of its uncertainty or less than 10\% of
the expected Standard Model signal yield over a wide range of Higgs boson mass hypotheses. The
functional form satisfying these criteria with the smallest number of free parameters is used to describe the
background shape in the fit of the data.
In the four high \ptt\ categories, an exponential function in mass is used. In the six other
categories, the exponential of a second-order polynomial in mass is used.

Table~\ref{table:hsg1_categories} summarizes the expected signal rate, mass resolution and
background in the ten categories for the 7~TeV and 8~TeV data samples. Small differences in mass
resolution arise from the differences in the effective constant term
measured with $\ztoee$ events and from the lower pile-up level in the 7~TeV data.

\begin{table*}
  \centering
  \caption{Summary of the expected number of signal events in the 105--160~GeV mass range $n_\text{sig}$, the FWHM of mass
    resolution, $\sigma_\text{eff}$ (half of the smallest range
  containing 68\% of the signal events),
    number of background events $b$ in the smallest mass window containing
    90\% of the signal ($\sigma_\text{eff90}$), and the ratio $s/b$ and $s/\sqrt{b}$ with $s$ the expected
    number of signal events in the window containing 90\% of signal events, for the $H\rightarrow\gamma\gamma$ channel. $b$ is derived from the fit of the data in the 105--160~GeV mass range.
    The value of $m_H$ is
    taken to be 126~GeV and the signal yield is assumed to be the expected Standard Model value. The
    estimates are shown separately for the 7~TeV and 8~TeV datasets and for the inclusive sample as well
    as for each of the categories used in the analysis. \label{table:hsg1_categories}}
  \vspace{0.1cm}
  \begin{small}
  \begin{tabular}{lcccccc}
\hline 
Category & $n_\text{sig}$ & FWHM [GeV] & $\sigma_\text{eff}$ [GeV] & $b$ in $\pm\sigma_\text{eff90}$ & $s/b$ [\%] & $s/\sqrt{b}$ \\ 
\hline
 \multicolumn{7}{c}{$\sqrt{s}$=8~TeV} \\
\hline
 Inclusive                   & 402.   & 3.69 & 1.67 & 10670 & 3.39  & 3.50 \\
 Unconv. central low $\ptt$  & 59.3   & 3.13 & 1.35 & 801   & 6.66  & 1.88 \\
 Unconv. central high $\ptt$ & 7.1    & 2.81 & 1.21 & 26.0  & 24.6  & 1.26 \\
 Unconv. rest low $\ptt$     & 96.2   & 3.49 & 1.53 & 2624  & 3.30  & 1.69 \\
 Unconv. rest high $\ptt$    & 10.4   & 3.11 & 1.36 & 93.9  & 9.95  & 0.96 \\
 Unconv. transition          & 26.0   & 4.24 & 1.86 & 910   & 2.57  & 0.78 \\
 Conv. central low $\ptt$    & 37.2   & 3.47 & 1.52 & 589   & 5.69  & 1.38 \\
 Conv. central high $\ptt$   & 4.5    & 3.07 & 1.35 & 20.9  & 19.4  & 0.88 \\
 Conv. rest low $\ptt$       & 107.2  & 4.23 & 1.88 & 3834  & 2.52  & 1.56 \\
 Conv. rest high $\ptt$      & 11.9   & 3.71 & 1.64 & 144.2 &  7.44 & 0.89 \\
 Conv. transition            & 42.1   & 5.31 & 2.41 & 1977  & 1.92  & 0.85 \\
\hline
 \multicolumn{7}{c}{$\sqrt{s}$=7~TeV} \\
\hline
 Inclusive                   & 73.9   & 3.38 & 1.54 & 1752  & 3.80  & 1.59 \\
 Unconv. central low $\ptt$  & 10.8   & 2.89 & 1.24 & 128   & 7.55  & 0.85 \\
 Unconv. central high $\ptt$ & 1.2    & 2.59 & 1.11 & 3.7   & 30.0  & 0.58 \\
 Unconv. rest low $\ptt$     & 16.5   & 3.09 & 1.35 & 363   & 4.08  & 0.78 \\
 Unconv. rest high $\ptt$    & 1.8    & 2.78 & 1.21 & 13.6  & 11.6  & 0.43 \\
 Unconv. transition          & 4.5    & 3.65 & 1.61 & 125   & 3.21  & 0.36 \\
 Conv. central low $\ptt$    & 7.1    & 3.28 & 1.44 & 105   & 6.06  & 0.62 \\
 Conv. central high $\ptt$   & 0.8    & 2.87 & 1.25 & 3.5   & 21.6  & 0.40 \\
 Conv. rest low $\ptt$       & 21.0   & 3.93 & 1.75 & 695   & 2.72  & 0.72 \\
 Conv. rest high $\ptt$      & 2.2    & 3.43 & 1.51 & 24.7  & 7.98  & 0.40 \\
 Conv. transition            & 8.1    & 4.81 & 2.23 & 365   & 2.00  & 0.38 \\
\hline
\end{tabular}
\end{small}

\end{table*}

\subsection{Mass measurement method}
\label{sec:hsg1_massmeas}

The mass spectra for the ten data categories and the two center-of-mass energies are fitted
simultaneously assuming the signal-plus-background hypothesis, using an unbinned maximum likelihood
fit with background and signal parameterization described in the previous sections.  
The fitted parameters of interest for the signal are the Higgs boson mass and the signal strength, defined as
the yield normalized to the SM prediction. The parameters describing the background mass
distributions for each category and center-of-mass energy are also free in the fit. The
systematic uncertainties are described by a set of nuisance parameters in the likelihood.  They
include uncertainties affecting the signal mass peak position, modeled as Gaussian constraints,
uncertainties affecting the signal mass resolution and uncertainties affecting the signal yield.

Figure~\ref{fig:hsg1_dataFits} shows the result of the simultaneous fit to the data over all
categories. For illustration, all categories are summed together, with a weight given by the signal-to-background ($s/b$)
ratio in each category.

\begin{figure}[hptb!]
  \centering
  \includegraphics[width=\singlePlotSize]{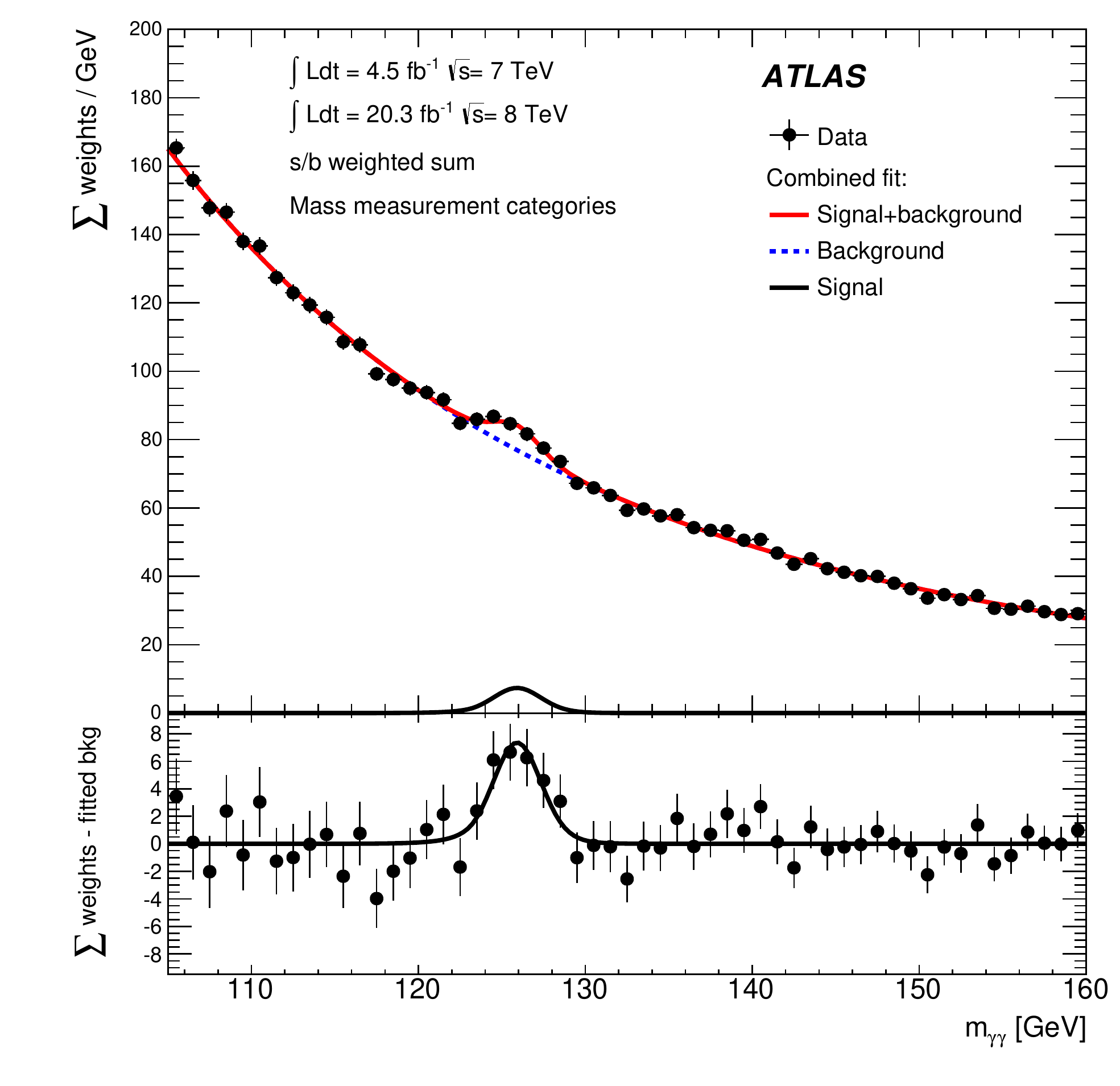}
  \caption{Invariant mass distribution in the \hgg\ analysis for data (7~TeV and 8~TeV samples
    combined), showing weighted data points with errors, and the result of the simultaneous fit to
    all categories.  The fitted signal plus background is shown, along with the background-only
    component of this fit.  The different categories are summed together with a weight given by the
    $s/b$ ratio in each category.  The bottom plot shows the
    difference between the summed weights and the background component of the fit.}\label{fig:hsg1_dataFits}
\end{figure}

\subsection{Systematic uncertainties}
\label{sec:hsg1_syst}

The dominant systematic uncertainties on the mass measurement arise from uncertainties on the photon
energy scale.  These uncertainties, discussed in Sec.~\ref{sec:egamma_sys}, are propagated to the
diphoton mass measurement in each of the ten categories. The total uncertainty on the mass
measurement from the photon energy scale uncertainties ranges from 0.17\% to 0.57\% depending on the
category. The category with the lowest systematic uncertainty is the low \ptt\ {\it central}
converted category, for which the energy scale extrapolation from $\ztoee$ events is the smallest.

Systematic uncertainties related to the reconstruction of the diphoton primary vertex are
investigated using $\ztoee$ events reweighted to match the transverse momentum distribution
of the Higgs boson and the $\eta$ distribution of the decay products.
The primary vertex is reconstructed using the same technique as
for diphoton events, ignoring the tracks associated with the electrons, and treating them as
unconverted photons. The dielectron invariant mass is then computed in the same way as the diphoton
invariant mass. Comparing the results of this procedure in data and simulation leads to an
uncertainty of 0.03\% on the position of the peak of the reconstructed invariant mass.

Systematic uncertainties related to the modeling of the background are estimated by performing
signal-plus-background fits to samples containing large numbers of simulated background events plus
the expected signal at various assumed Higgs boson masses. The signal is injected using the same
functional form used in the fit, so the fitted Higgs boson mass is sensitive only to the accuracy of
the background modeling. The maximum difference between the fitted Higgs boson mass and the input
mass over the tested mass range is assigned as a systematic uncertainty on the mass
measurement. This uncertainty varies from 0.05\% to 0.20\% depending on the category.  The
uncertainties in the different categories are taken as uncorrelated.  As a cross-check, to
investigate the impact of a background shape in data different than in the large statistics
simulated background sample, signal-plus-background pseudo-experiments are generated using a
functional form for the background with one more degree of freedom than the nominal background model
used in the fit: for the four high \ptt\ categories, a second-order Bernstein polynomial or the exponential
of a second-order polynomial is used; for the six other categories, a third-order Bernstein polynomial is used.
The parameters of the functional form used to generate these pseudo-experiments
are determined from the data.  These pseudo-experiments are then fitted using the nominal background
model.  This procedure leads to an uncertainty on the mass measurement between 0.01\% and 0.05\%
depending on the category, and smaller than the uncertainties derived from the baseline method using the
large sample of simulated background events.

Systematic uncertainties on the diphoton mass resolution due to uncertainties on the energy
resolution vary between 9\% and 16\% depending on the category and have a negligible impact on the
mass measurement.

Systematic uncertainties affecting the relative signal yield in each category arise from
uncertainties on the photon conversion rate, uncertainties in the proper classification of converted
and unconverted photon candidates and uncertainties in the modeling of the transverse momentum of
the Higgs boson. These migration systematic uncertainties vary between 3\% for the low \ptt\ categories,
dominated by uncertainties on the efficiency for reconstructing photon conversions, and 24\% for the
gluon fusion production process in the high \ptt\ categories, dominated by the uncertainty on the
transverse momentum of the Higgs boson. The uncertainty on the transverse momentum of the Higgs
boson is estimated by changing the renormalization and factorization scales in
the HRes2~\cite{deFlorian:2012mx, Grazzini:2013mca} computation of the Higgs boson transverse momentum distribution
as well as the resummation scales associated with
$t$- and $b$-quarks. These migration uncertainties have a negligible effect on the mass measurement.

Finally, uncertainties on the predicted overall signal yield are estimated as follows~\cite{ATLASggFinal}.
The uncertainty on the predicted cross-section for Higgs boson
production is about 10\% for the dominant gluon fusion process.  The uncertainty on the predicted
branching ratio to two photons is 5\%.  The uncertainty from the photon identification efficiency is
derived from studies using several control samples: a sample of radiative $Z$ decays, a sample of
$\ztoee$ events, where the shower shapes of electrons are corrected to resemble the shower shapes of photons,
and a sample of high \et\ isolated
prompt photons.  The estimated photon identification uncertainty amounts to 1.0\% for the 8~TeV
dataset, after correcting for small residual differences between simulation and data,
and 8.4\% for the 7~TeV dataset. 
The uncertainty is larger for the 7~TeV dataset because of the stronger 
correlation of the neural network photon identification with the photon isolation, and because the neural
network identification
relies more strongly on the correlations between the individual shower shape 
variables, complicating the measurement and introducing larger uncertainties 
on the estimate of its performance in data.
The uncertainty on the integrated luminosity is 2.8\% for
the 8~TeV dataset and 1.8\% for the 7~TeV dataset~\cite{ref:AtlasLuminosity-2011-final}.  
The uncertainties on the isolation cut efficiency and on the trigger efficiency are less than 1\%
for both the 7~TeV and 8~TeV datasets. These uncertainties on the overall signal
yield also have a negligible effect on the mass measurement.

Table~\ref{table:hsg1_sys} gives a summary of the systematic uncertainties on the mass measurement
for the different categories. For illustration, the 29 sources of uncertainty on the photon energy
scale are grouped into seven classes, so the correlations in the uncertainties per class
between categories are not 100\%.

\begin{table*}
\footnotesize
\centering
\caption{Summary of the relative systematic uncertainties (in \%) on the \hgg\ mass measurement for
  the different categories described in the text.  The first seven rows give the impact of the
  photon energy scale systematic uncertainties, grouped into seven classes. \label{table:hsg1_sys}}
  \vspace{0.1cm}
  \begin{tabular}{l|cccccccccc}
    \hline
  &  \multicolumn{5}{c}{Unconverted} & \multicolumn{5}{c}{Converted} \\
  &  \multicolumn{2}{c}{Central} & \multicolumn{2}{c}{Rest} & Trans. & \multicolumn{2}{c}{Central} & \multicolumn{2}{c}{Rest} & Trans. \\
 Class                        &  low $\ptt$ & high $\ptt$ & low $\ptt$ & high $\ptt$ & & low $\ptt$ & high $\ptt$ & low $\ptt$ & high $\ptt$ & \\
 \hline
$\ztoee$ calibration          &    0.02 &    0.03 &    0.04 &    0.04 &    0.11 &    0.02 &    0.02 &    0.05 &    0.05 &    0.11 \\
LAr cell non-linearity        &    0.12 &    0.19 &    0.09 &    0.16 &    0.39 &    0.09 &    0.19 &    0.06 &    0.14 &    0.29\\
Layer calibration             &    0.13 &    0.16 &    0.11 &    0.13 &    0.13 &    0.07 &    0.10 &    0.05 &    0.07 &    0.07 \\
ID material                   &    0.06 &    0.06 &    0.08 &    0.08 &    0.10 &    0.05 &    0.05 &    0.06 &    0.06 &    0.06 \\
Other material                &    0.07 &    0.08 &    0.14 &    0.15 &    0.35 &    0.04 &    0.04 &    0.07 &    0.08 &    0.20 \\
Conversion reconstruction     &    0.02 &    0.02 &    0.03 &    0.03 &    0.05 &    0.03 &    0.02 &    0.05 &    0.04 &    0.06 \\
Lateral shower shape          &    0.04 &    0.04 &    0.07 &    0.07 &    0.06 &    0.09 &    0.09 &    0.18 &    0.19 &    0.16 \\
\hline
Background modeling           &  0.10 &    0.06 &    0.05 &    0.11 &    0.16 &    0.13 &    0.06 &    0.14 &    0.18 &    0.20 \\
\hline
Vertex measurement            &  \multicolumn{10}{c}{ 0.03}                                                                     \\
\hline
Total                         &    0.23 &    0.28 &    0.24 &    0.30 &    0.59 &    0.21 &    0.25 &    0.27 &    0.33 &    0.47 \\
 \hline
\end{tabular}
\end{table*}

The total systematic uncertainty on the measured mass is $\pm$0.22\%, dominated by the
uncertainty on the photon energy scale.

\subsection{Result}
\label{sec:hsg1_res}

The measured Higgs boson mass in the $H\rightarrow\gamma\gamma$ decay channel is:
%
  \begin{equation}
    \begin{aligned}
      m_{H}  &= 125.98 \pm 0.42 {\rm{(stat)}} \pm 0.28 {\rm{(syst)}}  \rm{~GeV}\\  &= 125.98 \pm 0.50 \gev 
    \end{aligned}
  \end{equation}
where the first error represents the statistical uncertainty and the second the systematic
uncertainty. 
The change in central value compared to the previous result in
Ref.~\cite{ATLAScouplings} of $126.8\pm0.2{\rm{(stat)}} \pm 0.7{\rm{(syst)}}  \rm{~GeV}$ 
is consistent with the expected change resulting from the
updated photon energy scale calibration and its much smaller systematic uncertainty. 
From the changes in the calibration procedure an average shift of
about $-0.45$~GeV in the measured Higgs boson mass is expected, with
an expected statistical spread of about 0.35~GeV from fluctuations in
the measured masses of individual events. The average shift between
the old and new calibrations is estimated from the distribution of the mass difference of
the common events in the mass sidebands outside the signal region.

The mass measurement is performed
leaving the overall signal strength free in the fit. The measured signal strength, $\mu$, normalized
to the Standard Model expectation is found to be $\mu=1.29\pm0.30$. 
The most precise results for $\mu$ from this data are based on an analysis optimized to
measure the signal strength~\cite{ATLASggFinal}.
The statistical uncertainties
on the mass and signal yield obtained from the data fit are consistent with the expected statistical
accuracy in pseudo-experiments generated with this measured signal yield. The average expected
statistical uncertainty on the mass for $\mu=1.3$ is 0.35~GeV and the fraction of pseudo-experiments with a
statistical error larger than the one observed in data (0.42~GeV) is about 16\%. From these
pseudo-experiments, the distribution of fitted masses is compared to
the input mass value to verify that the average statistical uncertainty provides 68\% coverage.
In the previous measurement, the expected statistical uncertainty was about 0.33~GeV for $\mu=1.55$
and the observed statistical uncertainty (0.24~GeV) was better than expected.
The change in expected statistical uncertainty mostly comes from the change in the fitted signal strength, which was slightly
larger in the previous measurement, as the
statistical uncertainty on the mass measurement is inversely proportional to the signal strength.
Changes in the mass resolution and the event categorization also contribute to the change in the expected statistical uncertainty.
The increase in the statistical uncertainty between the previous result and this result is consistent with
a statistical fluctuation from changes in the measured masses of individual events.
Assuming the SM signal yield ($\mu=1$), the
statistical uncertainty on the mass measurement is expected to be 0.45~GeV.  

No significant shift in the values of the nuisance parameters
associated with the systematic uncertainties is observed in the fit to
the data. The result is also stable if a different mass range, 115~GeV
to 135~GeV, is used in the fit.

Several cross-checks of the mass measurement are performed, dividing the data into subsamples with
different sensitivities to systematic uncertainties. To evaluate the compatibility between the mass
measured in a subsample and the combined mass from all other subsamples, a procedure similar to the one used
to evaluate the mass compatibility between different channels, described in
Sec.~\ref{sec:stat_and_sys}, is applied.  The mass difference $\Delta_i$ between the subsample
$i$ under test and the combined mass is added as a parameter in the likelihood, and the value of
$\Delta_i$ with its uncertainty is extracted from the fit to the data, leaving the combined Higgs
boson mass as a free parameter. With this procedure, the uncertainty on $\Delta_i$ correctly
accounts for the correlation in systematic uncertainties between the subsample under test and the
rest of the dataset.  The values of $\Delta_i$ with their uncertainties are shown in
Fig.~\ref{fig:hsg1_delta} for three different alternative event categorizations, with three
subsamples each: as a function of the conversion status of the two photons, as a function of the
number of primary vertices reconstructed in the event and as a function of the photon impact point
in the calorimeter (barrel vs end-cap).  No value of $\Delta_i$ inconsistent with zero is found in
these checks, or in other categorizations related to the conversion topology, the instantaneous
luminosity, the photon isolation and the data taking periods.  A similar procedure, fitting
simultaneously one $\Delta_i$ per subsample, is performed to assess the global consistency of all
the different subsamples with a common combined mass. In nine different categorizations, no global
inconsistency larger than 1.5$\sigma$ is observed.

\begin{figure}[hptb!]
  \centering
  \includegraphics[width=\singlePlotSize]{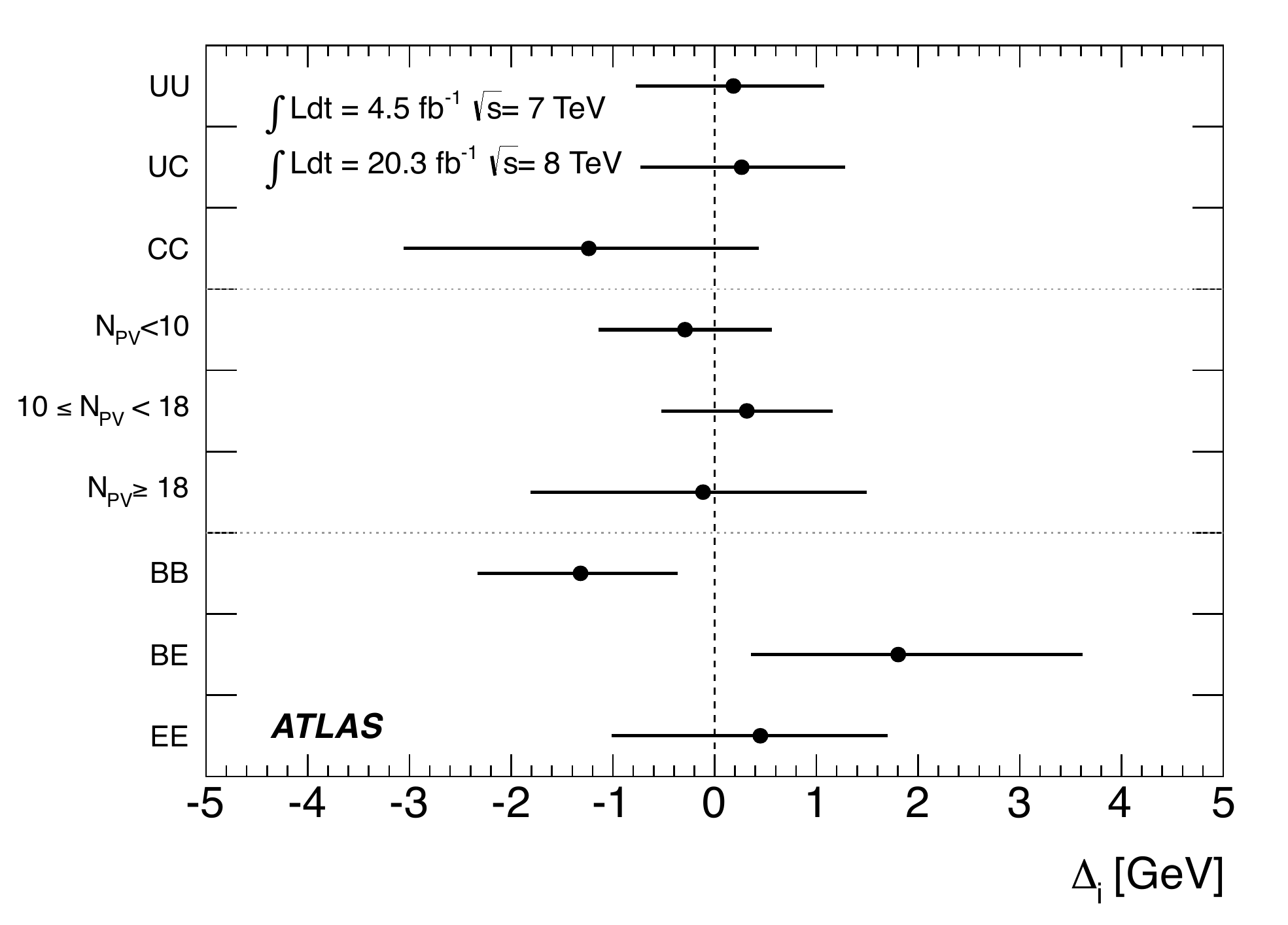}
  \caption{Difference, $\Delta_i$, between the mass measured in a given $\gamma\gamma$ subsample
    and the combined $\gamma\gamma$ mass, using three different alternative categorizations to
    define the subsamples. The top three points show a categorization based on the photon
    conversion status: $UU$ is the subsample with both photons unconverted, $UC$ the subsample
    with one converted and one unconverted photon, $CC$ the subsample with two converted
    photons. The middle three points show a categorization based on the number of reconstructed
    primary vertices ($N_{PV}$) in the event. The bottom three points show a categorization based on
    the photon impact points on the calorimeter: $BB$ is the subsample with both photons detected
    in the barrel calorimeter, $BE$ the subsample with one photon in the barrel calorimeter and one
    photon in the end-cap calorimeter and $EE$ the subsample with both photons in the end-cap
    calorimeter.  } \label{fig:hsg1_delta}
\end{figure}

A direct limit on the decay width of the Higgs boson is set from the observed width of the invariant
mass peak,  under the assumption that there is no interference with background processes.
The signal model is extended by convolving the detector resolution with
a non relativistic Breit--Wigner distribution to model a non-zero decay width.  
The test statistic used to obtain the limit on the width is  a
profile likelihood estimator with the width as main parameter of interest, where also the mass and the signal strength of the 
observed particle are treated as free parameters. 
Pseudo-experiments with different assumed widths are performed to estimate the distribution of the test statistic, which
does not perfectly follow a $\chi^2$ distribution, and
to compute the exclusion level. 
The observed (expected for
$\mu=1$) 95\% confidence level (CL) upper limit on the width is 5.0 (6.2) GeV.
These limits, properly calibrated with pseudo-experiments, are about 15\% different from estimates based on
a $\chi^2$ distribution of the test statistics.

\clearpage  
\section{Mass and width measurement in the \hzzllll\ channel \label{sec:hsg2}}

The \hzzllll\ channel provides good sensitivity to the measurement of the Higgs properties due to
its high signal-to-background ratio, which is about two in the signal mass window 120--130 GeV, and
its excellent mass resolution, for each of the four final states: $\mu^+\mu^-\mu^+\mu^-$~($4\mu$),
$e^+e^-\mu^+\mu^-$ ($2e2\mu$), $\mu^+\mu^-e^+e^-$ ($2\mu 2e$), and $e^+e^-e^+e^-$~($4e$), where the
first pair is defined to be the one with the dilepton mass closest to the $Z$ boson mass.  The
typical mass resolution varies from 1.6 GeV for the $4 \mu$ final state to 2.2 GeV for the $4 e$
final state.  For a SM Higgs boson with a mass of about 125 GeV, the dominant background is the
$(Z^{(*)}/\gamma^{*}) (Z^{(*)}/\gamma^{*})\rightarrow 4\ell$ process, referred to hereafter as
\ZZbkg.  A smaller contribution is expected from the $Z+\rm{jets}$ and $t\bar{t}$ processes.

Several improvements were introduced in the analysis with respect to Ref.~\cite{ATLAScouplings}.
For the 8 TeV data, the electron identification was changed from a cut-based to a likelihood method,
which improves the rejection of light-flavor jets and photon conversions by a factor of two for the
same signal efficiency~\cite{ElectronEff2012}.  The updated electromagnetic calibration based on
multivariate techniques, described in Sec.~\ref{sec:egamma_sys}, is used for electrons and
final-state radiation (FSR) photons.  In addition, a new combined fit of the track momentum and
cluster energy was introduced.  This is applied to electrons with \et\ $<$ 30 GeV when the track
momentum and cluster energy are consistent within their uncertainties, and improves the resolution
of the $m_{4\ell}$ invariant mass distribution for the $H\rightarrow ZZ^{*}\rightarrow 4e$ and
$H\rightarrow ZZ^{*}\rightarrow 2\mu 2e$ final states by about 4\%.  Finally, a multivariate
discriminant was introduced to separate the signal and \ZZbkg\ background.

The following subsections describe the details of the Higgs mass measurement in the
\hzzllll\ channel. A more complete discussion of the selection criteria and background determination
is reported in Ref.~\cite{ATLAS4lfinal}.

\subsection{Event selection}

Four-lepton events are selected with single-lepton and dilepton triggers.  The \pt\ (\et) thresholds
for single-muon (single-electron) triggers increased from 18 GeV to 24 GeV (20 GeV to 25 GeV)
between the 7 and 8 TeV datasets, due to the increase of the instantaneous luminosity during these
two data-taking periods. The dilepton triggers include dimuon, dielectron and mixed electron and
muon topologies, and have thresholds starting at 6 GeV (10 GeV) for muons (electrons) for 7 TeV
data.  For the 8 TeV data, the dilepton trigger thresholds were raised to 13 GeV for the dimuon and
to 12 GeV for the dielectron.  In addition for the 8 TeV data, an asymmetric threshold of (8,18) GeV
was added for the dimuon trigger.  The trigger efficiency for Higgs boson signal events passing the
final selection is greater than $97\%$ for the $4\mu$, $2e2\mu$ and $2\mu 2e$ channels and close to
100\% for the $4e$ channel.

For the 7 TeV data, electrons are required to satisfy a cut-based selection using tracking and
shower profile criteria~\cite{ElectronEff2011}.  The 8 TeV data have an improved electron
reconstruction algorithm with higher efficiency, and the likelihood-based electron identification
with improved background rejection mentioned above.  The four types of muons described in
Sec.~\ref{sec:muon_sys} are allowed with at most one {\it standalone} or {\it calorimeter-tagged}
muon per event.  Muon tracks are required to have a minimum number of hits in the ID, or hits in all
muon stations for {\it standalone} muons.

Higgs boson candidates are formed by selecting two same-flavor, opposite-sign lepton pairs (a lepton
quadruplet) in an event.  Each lepton is required to have a longitudinal impact parameter less than
10 mm with respect to the primary vertex, defined as the primary vertex with the largest $\sum
\pt^2$, and muons are required to have a transverse impact parameter less than 1 mm to reject
cosmic-ray muons.  Each muon (electron) must satisfy $\pt>6$ GeV ($\et>7$ GeV) and be measured in
the pseudorapidity range $\left|\eta\right|<2.7$ ($\left|\eta \right|<2.47$). The highest \pt~lepton
in the quadruplet must satisfy $\pt >20$ GeV, and the second (third) lepton in $\pt$ order must
satisfy $\pt>15$ GeV ($\pt>10$ GeV). The leptons are required to be separated from each other by
$\Delta R>$ 0.1 (0.2) for same (different) flavor.  Each event is required to have the triggering
lepton(s) matched to one or two of the selected leptons.

Multiple quadruplets within a single event are possible: for four muons or electrons there are two
ways to pair the leptons, and for five or more leptons there are multiple ways to choose the
leptons.  Quadruplet selection is done separately in each channel: $4\mu$, $2e2\mu$, $2\mu2e$, $4e$,
keeping only a single quadruplet per channel.  For each channel, the lepton pair with the mass
closest to the $Z$ boson mass is selected as the leading dilepton pair and its invariant mass,
$m_{12}$, is required to be between 50~GeV and 106~GeV.  The second, subleading, pair of each
channel is chosen as the pair with its invariant mass, $m_{34}$, closest to the $Z$ mass, and also
satisfying $m_{\rm{min}}< m_{34} < 115$ GeV. Here $m_{\rm{min}}$ takes the value of 12 GeV for
$m_{4\ell}<$ 140 GeV, increases linearly between 12 and 50 GeV for $140 < m_{34} < 190$ GeV, and is
50 GeV for $m_{4\ell} > $ 190 GeV.  Finally, if the event contains a quadruplet passing the
selection in more than one channel, the quadruplet from the channel with the highest expected rate
is taken, i.e. the first is taken from the order: $4\mu$, $2e2\mu$, $2\mu2e$, $4e$.

The $Z+\rm{jets}$ and $t\bar{t}$~background contributions are further reduced by applying impact
parameter and track- and calorimeter-based isolation requirements to the leptons. The impact
parameter significance, $|d_0|/\sigma_{d_0}$, for all muons (electrons) is required to be less than
3.5 (6.5).  The normalized track isolation discriminant, defined as the sum of the transverse
momenta of tracks inside a cone of size $\Delta R=0.2$ around the lepton, excluding the lepton
track, divided by the lepton \pt, is required to be smaller than 0.15.  The normalized calorimetric
isolation is computed from the energy in the electromagnetic and hadronic calorimeters within a cone
of $\Delta R<0.2$ around the lepton, excluding the cells containing the lepton energy.  This energy
is corrected, event-by-event, for the ambient energy deposition in the event from pile-up as well as
for the underlying event, and then divided by the lepton \pt.  The normalized calorimetric isolation
is required to be smaller than 0.2 (0.3) for electrons in the 7 TeV (8 TeV) data, and smaller than
0.3 for muons (0.15 for {\it standalone} muons).

The effect of photon emission from final-state radiation (FSR) on the reconstructed invariant mass
is well modeled in the simulation.  In addition, some FSR recovery is performed allowing at most one
photon to be added per event.  Leading dimuon candidates with $m_{12}$ in the range 66--89 GeV,
below the $Z$ boson mass, are corrected for collinear FSR by including in the invariant mass any
reconstructed photon lying close to one of the muon tracks, as long as the corrected mass,
$m_{\mu\mu\gamma}$, remains below 100 GeV.  In a second step, for events without collinear FSR,
non-collinear FSR photons with a significant \et\ are included for both the leading dimuon and
dielectron candidates, an improvement introduced since Ref.~\cite{ATLAScouplings}.  The expected
number of events with a collinear or non-collinear FSR correction is 4\% and 1\%, respectively.
Full details are discussed in Ref.~\cite{ATLAS4lfinal}.

For the 8 TeV data, the combined signal reconstruction and selection efficiency for $\mH=125$ GeV is
$39$\% for the $4\mu$ channel, $27$\% for the $2e2\mu$/$2\mu2e$ channel and $20$\% for the $4e$
channel.

Finally, a kinematic fit is used to constrain the mass of the leading lepton pair to the $Z$ pole
mass within the experimental resolution, including any FSR photon, as in the analysis of
Ref.~\cite{ATLAScouplings}.  This improves the $m_{4\ell}$ resolution by about 15\%.

\subsection{Background estimation}

The \ZZbkg\ background is estimated from simulation and normalized to NLO
calculations~\cite{Dittmaier:2012vm}.  The reducible $Z+\rm{jets}$ and $t\bar{t}$ backgrounds are
estimated with data-driven methods, separately for the two final states with subleading muons,
$\ell\ell+\mu\mu$, and the two final states with subleading electrons, $\ell\ell+ee$.  For the
$\ell\ell+\mu\mu$ reducible background, the $Z+{\rm jets}$ background mostly consists of
$Z+b\bar{b}$ events with heavy-flavor semileptonic decays and to a lesser extent
$\pi$/\kaon~in-flight decays.  The $Z+\rm{jets}$ and $t\bar{t}$ backgrounds can be distinguished in
the $m_{12}$ distribution where the former background peaks at the $Z$ boson mass, and the latter
has a broad distribution.  Four control regions, with relaxed impact parameter and isolation
selection on the subleading muons, are fit simultaneously to extract the different components of the
reducible background.  The four control regions are defined by: at least one subleading muon with
inverted impact parameter significance to enhance the heavy flavor contribution, at least one
subleading muon with inverted isolation significance to enhance the $\pi$/\kaon~in-flight decays,
same-sign subleading muons to include all contributions, and finally a leading $e\mu$ pair with
either a same- or an opposite-sign subleading muon pair, which removes the $Z+\rm{jets}$
contribution.  The fitted yields in the control regions are extrapolated to the signal region using
efficiencies obtained from simulation.  A small contribution from $WZ$ decays is estimated using
simulation.

The electron background contributing to the $\ell\ell+ee$ final states arises mainly from jets
misidentified as electrons, occurring in three ways: light-flavor hadrons misidentified as
electrons, photon conversions reconstructed as electrons, and non-isolated electrons from
heavy-flavor hadronic decays.  The electron background is evaluated by three data-driven methods
where the selection is relaxed or inverted for one or two of the subleading electrons.  The final
estimate is obtained using a ``$3\ell+X$'' control region, and the other methods, which are used as
cross-checks, are described in Ref.~\cite{ATLAS4lfinal}.  The $3\ell+X$ control region requires the
three highest \pt\ leptons ($3\ell$) to satisfy the full selection, with the third $\ell$ an
electron, and the remaining electron ($X$) to have the electron identification fully relaxed except
for the requirement on the number of hits in the silicon tracker -- at least seven silicon hits with
at least one in the pixel detector. In addition, the $X$ is required to have the same sign as the
other subleading electron to minimize the contribution from the \ZZbkg\ background.  The yields of
the background components of $X$ are extracted with a fit to the number of hits in the first pixel
layer (B-layer) and the high-threshold to low-threshold TRT hit ratio. Most photons have no B-layer
hit, and the TRT threshold distinguishes between the hadrons misidentified as electrons and the
photon-conversion and heavy-flavor electrons.  The fitted yields in the control region are
extrapolated to the signal region using efficiencies obtained from a large sample of $Z$ bosons
produced with a single additional electron candidate satisfying the relaxed selection.

To evaluate the background in the signal region, the $m_{4\ell}$ shape is evaluated using simulated
events for the $\ell\ell+\mu\mu$ final states and with data using the $3\ell+X$ method for the
$\ell\ell+ee$ final states.  The estimates for the \ZZbkg\ and reducible backgrounds in the $120 <
m_{4\ell} < 130$ GeV mass window are provided in Table~\ref{tab:yields}.

\subsection{Multivariate discriminant}
    
The multivariate discriminant used to reduce the impact of the \ZZbkg\ background on the fitted mass
is based on a boosted decision tree (BDT)~\cite{Hocker:2007ht}.  The BDT classifier
($\mathrm{BDT}_{\ZZbkg}$) is trained using simulated signal events generated with \mH\ $=125$ GeV
and simulated \ZZbkg\ background events that pass the event selection and have $115<
m_{4\ell}<130$~GeV, the mass window that contains over 95\% of the signal.  The variables used in
the training are the transverse momentum and the pseudorapidity of the four-lepton system, plus a
matrix-element-based kinematic discriminant ($D_{\ZZbkg}$) defined as:
\begin{equation}
  D_{\ZZbkg}=\ln\left(\frac{\left|{\cal{M}}_{\rm sig}\right|^{2}}{\left|{\cal{M}}_{ZZ}\right|^{2}}\right) ,
\end{equation}
\vspace{0.2cm}
where ${\cal{M}}_{\rm sig}$ and ${\cal{M}}_{ZZ}$ are the matrix elements for the signal and
\ZZbkg\ background processes, respectively, computed at leading order using
MadGraph~\cite{Alwall:2014hca}.

\subsection{Signal and background model}

Several methods are used to measure the Higgs boson mass in the \hzzllll\ decay channel.  The
two-dimensional (2D) fit to the $m_{4\ell}$ and $\mathrm{BDT}_{\ZZbkg}$ output ($O_{BDT_{\ZZbkg}}$)
is chosen as the baseline because it has the smallest expected uncertainty among the different
methods.  The one-dimensional (1D) fit to the $m_{4\ell}$ spectra used for the previous
measurement~\cite{ATLAScouplings} serves as a cross-check.  For both the 1D and 2D fits, the signal
model is based on simulation distributions that are smoothed using a kernel density estimation
method ~\cite{Cranmer:2000du}.  These distributions are generated at 15 different $m_{H}$ values in
the range $115<m_{H}<130$~GeV and form templates that are parameterized as a function of $m_{H}$
using B-spline interpolation~\cite{NURBS:1997}.  These simulation samples at different masses are
normalized to the expected SM cross-section times branching ratio~\cite{Dittmaier:2013} to derive
the expected signal yields after acceptance and selection.  For all of the methods, the $m_{4\ell}$
range used for the fit is 110~GeV to 140~GeV.

The signal probability density function (PDF) in the 2D fit is modeled as:  
\iftoggle{isPRD} {
  \begin{widetext}
    \begin{equation}
      \begin{aligned}
        \mathcal{P}(m_{4\ell},O_{BDT_{\ZZbkg}}~|~m_{H})&=\mathcal{P}(m_{4\ell}~|~O_{BDT_{\ZZbkg}},~m_{H})~ \mathcal{P}(O_{BDT_{\ZZbkg}}~|~m_{H})\\
        &\simeq  \left( \sum_{n=1}^{4} \mathcal{P}_n(m_{4\ell}~|~m_{H}) \theta_n (O_{BDT_{\ZZbkg}}) \right) \mathcal{P}(O_{BDT_{\ZZbkg}}~|~m_{H}) \\
      \end{aligned} \label{eq:sigpdf}
    \end{equation}
  \end{widetext}
}{ 
  \begin{equation}
    \begin{aligned}
      \mathcal{P}(m_{4\ell},O_{BDT_{\ZZbkg}}~|~m_{H})&=\mathcal{P}(m_{4\ell}~|~O_{BDT_{\ZZbkg}},~m_{H})~ \mathcal{P}(O_{BDT_{\ZZbkg}}~|~m_{H})\\
      &\simeq  \left( \sum_{n=1}^{4} \mathcal{P}_n(m_{4\ell}~|~m_{H}) \theta_n (O_{BDT_{\ZZbkg}}) \right) \mathcal{P}(O_{BDT_{\ZZbkg}}~|~m_{H}) \\
    \end{aligned} \label{eq:sigpdf}
  \end{equation}
}
where $\theta_{n}$ defines four equal-sized bins for the value of the $\mathrm{BDT}_{\ZZbkg}$
output, and $\mathcal{P}_n$ represents the 1D PDF for $m_{4\ell}$ for the signal in the
corresponding $O_{BDT_{\ZZbkg}}$ bin.  The variation of the $m_{4\ell}$ shape within a single
$O_{BDT_{\ZZbkg}}$ bin is found to be negligible and studies indicate that the binning approximation
does not bias the mass measurement.  The background model, $\mathcal{P}_{\rm
  bkg}(m_{4\ell},O_{BDT_{\ZZbkg}})$, is described using a full 2D PDF that is derived from
simulation for the \ZZbkg\ background, and by using data-driven techniques for the reducible
background.  The 2D template fit method reduces the expected statistical error on the measured mass
with respect to the simple fit to the $m_{4\ell}$ spectra (1D method) by about 8\%.

Extensive studies were performed in order to validate the signal and background PDFs using a 2D fit
to fully simulated signal and background events normalized to the SM expectation.  No bias was found
between the input and resulting 2D fit values for the Higgs mass and signal strength, tested for
different $m_H$ values in the range 120~GeV to 130~GeV.  Different values for the parameter used to
control the amount of smoothing for both the signal and background PDFs were tested and no biases on
the fitted $m_{H}$ and signal strength were found.  An additional check for a possible bias due to a
small dependence of the $\mathrm{BDT}_{\ZZbkg}$ output on $m_{H}$ for the signal, included in
Eq.~(\ref{eq:sigpdf}), is performed by fitting a sample of background-only simulated data. No
dependence of the likelihood scan on $m_{H}$ was observed.

In addition to the 2D fit method, described above, and the 1D fit method used in
Ref.~\cite{ATLAScouplings}, a third approach is used.  This approach combines an analytic
description of the signal mass spectra with the $\mathrm{BDT}_{\ZZbkg}$ output and can be used both
for the mass measurement and to provide a direct limit on the width of the Higgs boson.  In this
method, the signal $m_{4\ell}$ PDF is computed event-by-event by convolving the estimated detector
response for each of the four leptons with the non relativistic Breit--Wigner function describing
the generated Higgs mass line-shape.  The advantage of this method is that the typical detector
response for each data candidate is taken into account in the signal modeling. This is referred to
as the per-event-error method.  In this fit the $Z$ mass constraint is not applied.  The muon and
electron response functions are modeled by the sum of two or three Gaussian distributions,
respectively, to provide a better description of the responses. This parameterization is performed
in bins of \eta\ and \pt.  These response functions are validated with several simulation samples
and with data.  One validation consists of comparing the $Z$ boson mass distribution measured in
collision data with the convolution of the generator-level $Z$ boson resonance with the detector
response, constructed using the single-lepton response. The ratio of the two distributions agrees to
better than 2\% for $Z\rightarrow\mu\mu$ and 5\% for $Z\rightarrow ee$.  In addition, the
per-event-error model is checked by fitting the four-lepton invariant mass from the $Z$ decay in the
$Z \to 4 \ell$ process. The fit results are in agreement with the world average values of the $Z$
boson mass and width~\cite{ALEPH:2005ab}.  The per-event-error fit is used both as a cross-check for
the mass measurement and as the baseline method to set an upper limit on the Higgs boson total width
$\Gamma_{H}$.

For the mass measurement, the $m_{4\ell}$ (and $O_{BDT_{\ZZbkg}}$) data distributions for eight sets
of events, one for each final state for the $7$~TeV and $8$~TeV data, are simultaneously fitted
using an unbinned maximum likelihood assuming the signal and background models described above.  
The backgrounds are set in the fit to their estimated values and the associated normalization and
shape uncertainties are treated as nuisance parameters, as discussed in Sec.~\ref{sec:stat_and_sys}.

\subsection{Systematic uncertainties}
The main sources of systematic uncertainties on the mass measurement are the electron energy scale
and the muon momentum scale.
The expected impact of these uncertainties on the mass measurement corresponds to about 60 MeV 
for both the $4e$ and the $4\mu$ channels, obtained from the 2D fit to simulation.
When all the final states are combined together this translates to an observed $\pm 0.03\%$
uncertainty on $m_{H}$ for each of the electron energy scale and the muon momentum scale.

Systematic uncertainties on the measurement of the inclusive signal rate are also included in the
model.  The uncertainty on the inclusive signal strength due to the identification and
reconstruction efficiency for muons and electrons is $\pm 2\%$.  The dominant theory systematic
uncertainties arise from QCD scale variations of the $gg\rightarrow H$ process ($\pm 7\%$), parton
distribution function variations ($\pm 6\%$) and the decay branching ratio ($\pm 4\%$).  The uncertainty
on the Higgs boson transverse momentum, evaluated as described in Sec.~\ref{sec:hsg1_syst}, has a
negligible impact on the mass and the inclusive signal rate measurements.  The uncertainty on the
integrated luminosity is given in Sec.~\ref{sec:hsg1_syst}, and has a negligible impact on the 
mass measurement.

\subsection{Results}

Figure~\ref{fig:M4l} shows the $m_{4\ell}$ distribution of the selected candidates for $7$~TeV and
$8$~TeV collision data along with the expected distributions for a signal with a mass of 124.5 GeV
and the \ZZbkg\ and reducible backgrounds. The expected signal is normalized to the measured signal
strength, given below. Figure~\ref{fig:BDTvsM4l} shows the $\mathrm{BDT}_{\ZZbkg}$ output versus
$m_{4\ell}$ for the selected candidates in the $m_{4\ell}$ range 110--140~GeV.  The compatibility of
the data with the expectations shown in Fig.~\ref{fig:BDTvsM4l} has been checked using
pseudo-experiments generated according to the expected two-dimensional distributions and good
agreement has been found.  Table~\ref{tab:yields} presents the observed and expected number of
events for $\sqrt{s}=7$ TeV and $\sqrt{s}=8$ TeV, in a mass window of 120--130~GeV, corresponding to
about $\pm 2 \sigma_{m_{4\ell}}$.

\begin{figure}[hptb!]
  \centering
  \subfigure[\label{fig:M4l}]{\includegraphics[width=\doublePlotSize]{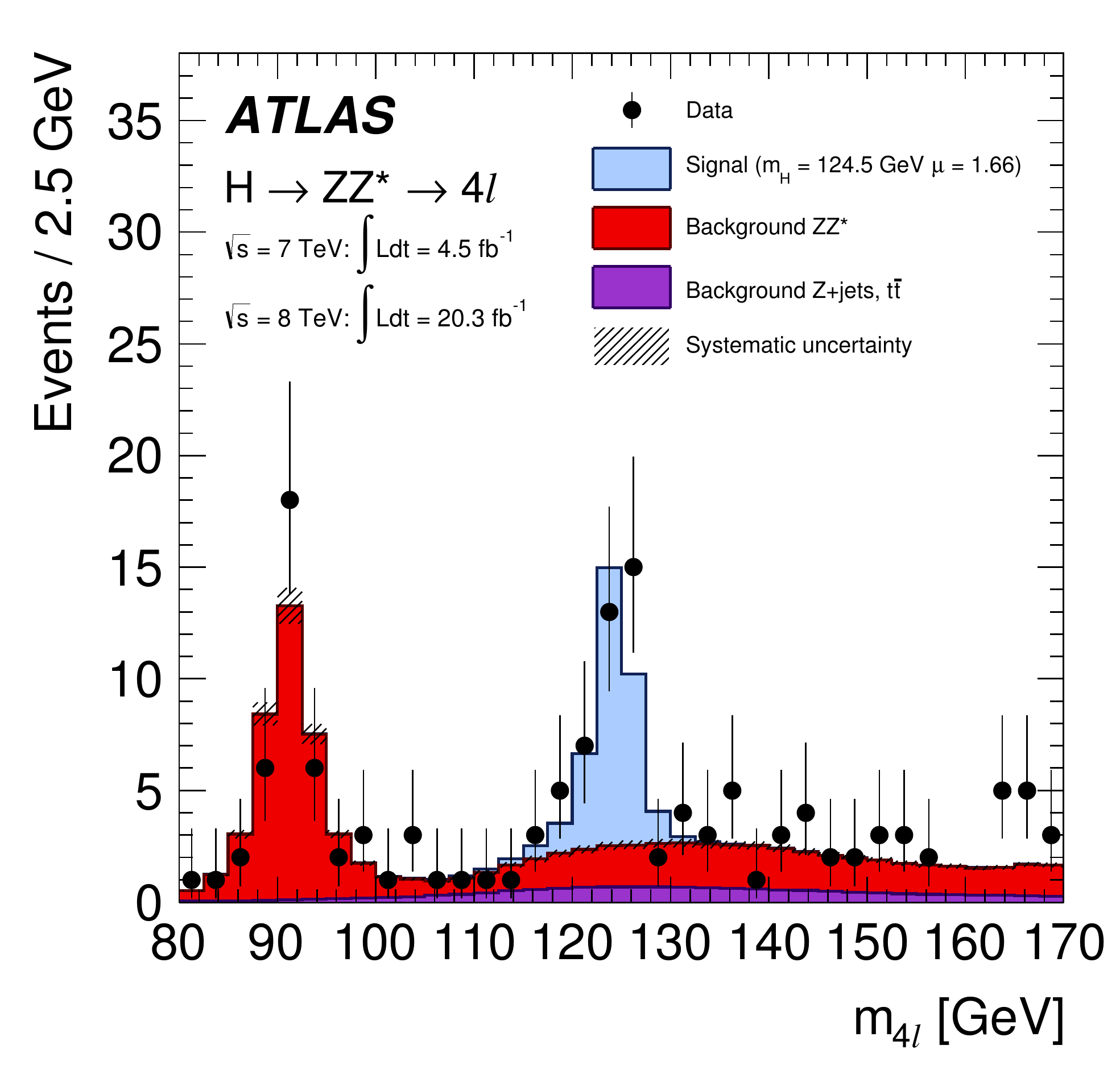}}
  \subfigure[\label{fig:BDTvsM4l}]{\includegraphics[width=\doublePlotSize]{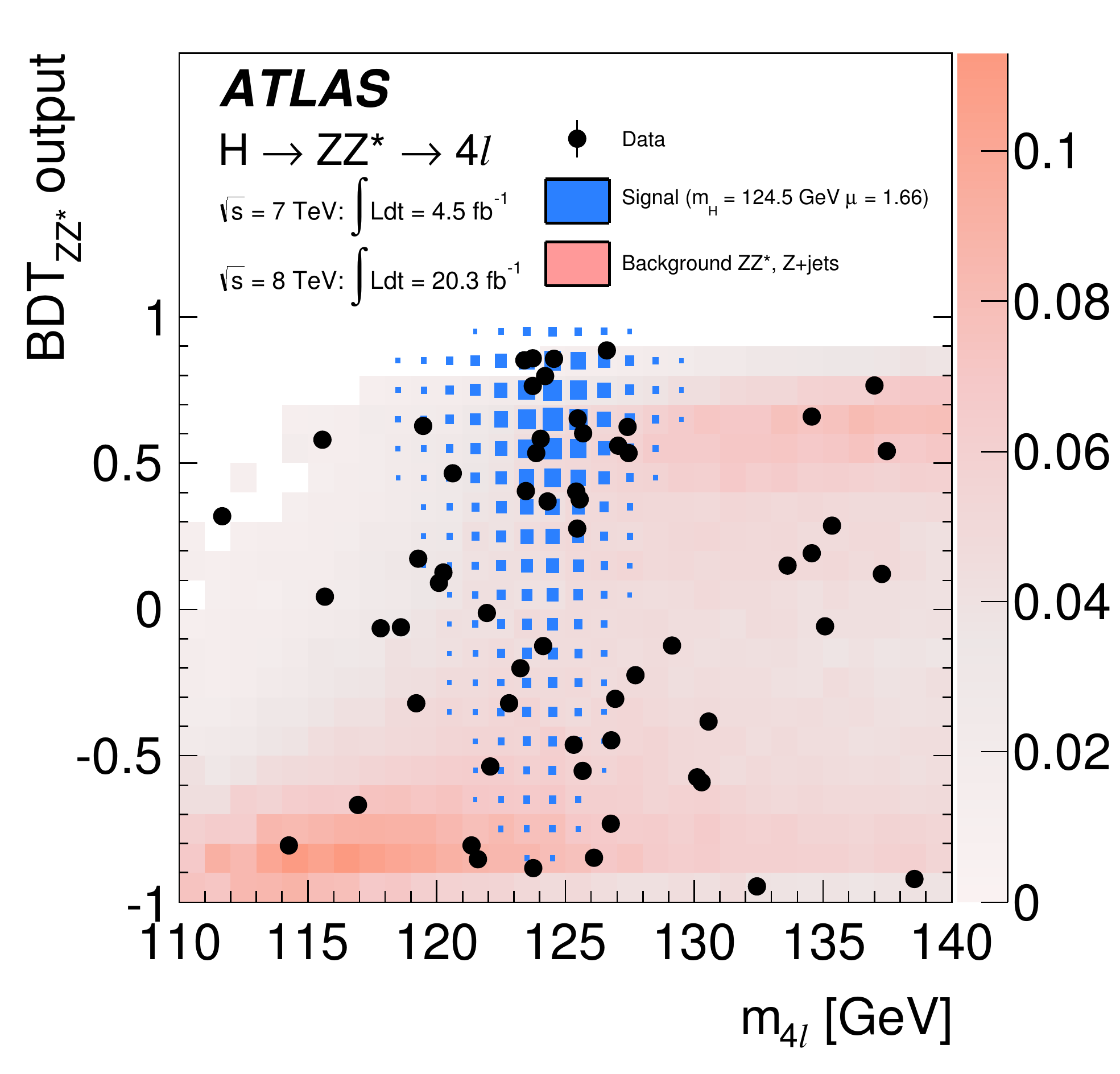}}
  \caption{~\subref{fig:M4l} Distribution of the four-lepton invariant mass for the selected
    candidates in the $m_{4\ell}$ range $80$--$170\,\gev$ for the combined $7$~TeV and
    $8$~TeV data samples. Superimposed are the expected distributions of a SM Higgs boson
    signal for $m_{H}$=124.5 GeV normalized to the measured signal strength, as well as the expected
    \ZZbkg\ and reducible backgrounds.  ~\subref{fig:BDTvsM4l} Distribution of the
    $\mathrm{BDT}_{\ZZbkg}$ output, versus $m_{4\ell}$ for the selected candidates in the
    $110$--$140\,\gev$ $m_{4\ell}$ range for the combined $7$~TeV and $8$ TeV data
    samples.  The expected distribution for a SM Higgs with $\mH=124.5\,\gev$ is indicated by the
    size of the blue boxes, and the total background is indicated by the intensity of the red
    shading. \label{fig:m4l}}
\end{figure}

\begin{table*}[htbb!]
  \centering 

  \caption{The number of events expected and observed for a $m_{H}$=$125$ GeV hypothesis for the
  four lepton final states.  The second column shows the number of expected signal events for the
  full mass range.  The other columns show the number of expected signal events, the number
  of \ZZbkg\ and reducible background events, the signal-to-background ratio ({\it s/b}), together
  with the numbers of observed events, in a window of $120 < m_{4\ell} < 130$ GeV for 4.5 \ifb\ at
  $\sqrt{s}=7$ TeV and 20.3 \ifb\ at $\sqrt{s}=8$ TeV as well as for the combined
  sample.  \label{tab:yields}}

  \setlength{\tabcolsep}{8pt}

  \vspace{0.1cm}
  \begin{tabular}{*{8}{c}}
      \hline\hline
    \noalign{\vspace{0.05cm}}


    Final state & Signal & Signal & \ZZbkg & $Z+\rm jets$,~$t\bar{t}$ & {\it s/b } & Expected & Observed \\
    & full mass range     \\

    \hline                                                     
    \noalign{\vspace{0.05cm}}
    \multicolumn{8}{c}{$\sqrt{s}=7$ TeV}\\
    \noalign{\vspace{0.05cm}}
    \hline
    $4\mu$    & 1.00 $\pm$  0.10  & 0.91 $\pm$  0.09 &  0.46 $\pm$  0.02 & 0.10 $\pm$  0.04 &  1.7 & 1.47 $\pm$  0.10 & 2 \\
    $2e2\mu$  & 0.66 $\pm$  0.06  & 0.58 $\pm$  0.06 &  0.32 $\pm$  0.02 & 0.09 $\pm$  0.03 &  1.5 & 0.99 $\pm$  0.07 & 2 \\
    $2\mu2e$  & 0.50 $\pm$  0.05  & 0.44 $\pm$  0.04 &  0.21 $\pm$  0.01 & 0.36 $\pm$  0.08 &  0.8 & 1.01 $\pm$  0.09 & 1 \\
    $4e$      & 0.46 $\pm$  0.05  & 0.39 $\pm$  0.04 &  0.19 $\pm$  0.01 & 0.40 $\pm$  0.09 &  0.7 & 0.98 $\pm$  0.10 & 1 \\
    \noalign{\vspace{0.05cm}}
    Total     & 2.62 $\pm$  0.26  & 2.32 $\pm$  0.23 &  1.17 $\pm$  0.06 & 0.96 $\pm$  0.18 &  1.1 & 4.45 $\pm$  0.30 & 6 \\

    \hline    					    
    \noalign{\vspace{0.05cm}}
    \multicolumn{8}{c}{$\sqrt{s}=8$ TeV}\\
    \noalign{\vspace{0.05cm}}
    \hline
    $4\mu$    & 5.80 $\pm$  0.57  &  5.28 $\pm$  0.52  & 2.36 $\pm$  0.12   & 0.69 $\pm$  0.13 &  1.7 &  8.33 $\pm$  0.6 & 12 \\
    $2e2\mu$  & 3.92 $\pm$  0.39  &  3.45 $\pm$  0.34  & 1.67 $\pm$  0.08   & 0.60 $\pm$  0.10 &  1.5 &  5.72 $\pm$  0.37 &  7 \\
    $2\mu2e$  & 3.06 $\pm$  0.31  &  2.71 $\pm$  0.28  & 1.17 $\pm$  0.07   & 0.36 $\pm$  0.08 &  1.8 &  4.23 $\pm$  0.30 &  5 \\
    $4e$      & 2.79 $\pm$  0.29  &  2.38 $\pm$  0.25  & 1.03 $\pm$  0.07   & 0.35 $\pm$  0.07 &  1.7 &  3.77 $\pm$  0.27 &  7 \\
    \noalign{\vspace{0.05cm}}
    Total     & 15.6 $\pm$  1.6    & 13.8  $\pm$  1.4   & 6.24 $\pm$  0.34   & 2.00 $\pm$  0.28 &  1.7 & 22.1 $\pm$  1.5 & 31 \\

    \hline
    \noalign{\vspace{0.05cm}}
    \multicolumn{8}{c}{$\sqrt{s}=7$ TeV and $\sqrt{s}=8$ TeV}\\
    \noalign{\vspace{0.05cm}}
    \hline
    $4\mu$    & 6.80 $\pm$  0.67  &  6.20 $\pm$  0.61  & 2.82 $\pm$  0.14  & 0.79 $\pm$  0.13 &  1.7 &  9.81 $\pm$  0.64 & 14 \\
    $2e2\mu$  & 4.58 $\pm$  0.45  &  4.04 $\pm$  0.40  & 1.99 $\pm$  0.10  & 0.69 $\pm$  0.11 &  1.5 &  6.72 $\pm$  0.42 &  9 \\
    $2\mu2e$  & 3.56 $\pm$  0.36  &  3.15 $\pm$  0.32  & 1.38 $\pm$  0.08  & 0.72 $\pm$  0.12 &  1.5 &  5.24 $\pm$  0.35 &  6 \\
    $4e$      & 3.25 $\pm$  0.34  &  2.77 $\pm$  0.29  & 1.22 $\pm$  0.08  & 0.76 $\pm$  0.11 &  1.4 &  4.75 $\pm$  0.32 &  8 \\
    \noalign{\vspace{0.05cm}}
    Total     &18.2  $\pm$  1.8   & 16.2  $\pm$  1.6   & 7.41 $\pm$  0.40  & 2.95 $\pm$  0.33 &  1.6 & 26.5  $\pm$  1.7 & 37 \\
    \hline\hline 
  \end{tabular}

\end{table*}

The measured Higgs boson mass in the \hzzllll\ decay channel obtained with the baseline 2D method is:
\begin{equation}
\begin{aligned}
        m_{H} &= 124.51 \pm 0.52 {\rm{~(stat)~}} \pm 0.06 {\rm{~(syst)~}} \gev \\ &= 124.51 \pm 0.52 \gev 
\end{aligned}
\end{equation}
where the first error represents the statistical uncertainty and the second the systematic
uncertainty.  
The systematic uncertainty is obtained from the quadrature subtraction of the fit uncertainty
evaluated with and without the systematic uncertainties fixed at their best fit values.  
Due to the large difference between the magnitude of the statistical and systematic uncertainties,
the numerical precision on the quadrature subtraction is estimated to be of the order of 10 MeV.
The measured signal strength for this inclusive selection is
$\mu=1.66^{+0.45}_{-0.38}$, consistent with the SM expectation of one.  The most precise results for
$\mu$ from this data are based on an analysis optimized to measure the signal
strength~\cite{ATLAS4lfinal}.  The expected statistical uncertainty for the 2D fit with the observed
$\mu$ value of 1.66 is 0.49 GeV, consistent with the observed statistical uncertainty.  With the
improved uncertainties on the electron and muon energy scales, the mass uncertainty given above is
predominantly statistical with a nearly negligible contribution from systematic uncertainties.  The
mass measurement performed with the 1D model gives $m_{H}=124.63 \pm 0.54$ GeV, consistent with the 2D
result where the expected difference has an RMS of 250 MeV estimated from Monte Carlo
pseudo-experiments.  These measurements can be compared to the previously reported
result~\cite{ATLAScouplings} of $124.3^{+0.6}_{-0.5} {\rm{~(stat)~}}^{+0.5}_{-0.3}{\rm{~(syst)}}$
GeV, which was obtained using the 1D model.  The difference between the measured values arises
primarily from the changes to the channels with electrons -- the new calibration and resolution
model, the introduction of the combined track momentum and cluster energy fit, and the improved
identification, as well as the recovery of non-collinear FSR photons, which affects all channels.
In the 120--130~GeV mass window, there are four new events and one missing event as compared to
Ref.~\cite{ATLAScouplings}.  Finally as a third cross-check, the measured mass obtained with the
per-event-error method is within 60 MeV of the value found with the 2D method.

Figure~\ref{fig:like4lscan} shows the scan of the profile likelihood, $-2\ln{\Lambda(m_{H})}$, for
the 2D model as a function of the mass of the Higgs boson for the four final states, as well as for
all of the channels combined.  The signal strength and all the nuisance parameters are profiled
(allowed to float to the values that maximize the likelihood) in the scan.  The compatibility among
the mass measurements from the four final states is estimated to be about 20\% using a $\chi^{2}$ test.

\begin{figure}[htbp!]
  \centering
  \includegraphics[width=\singlePlotSize]{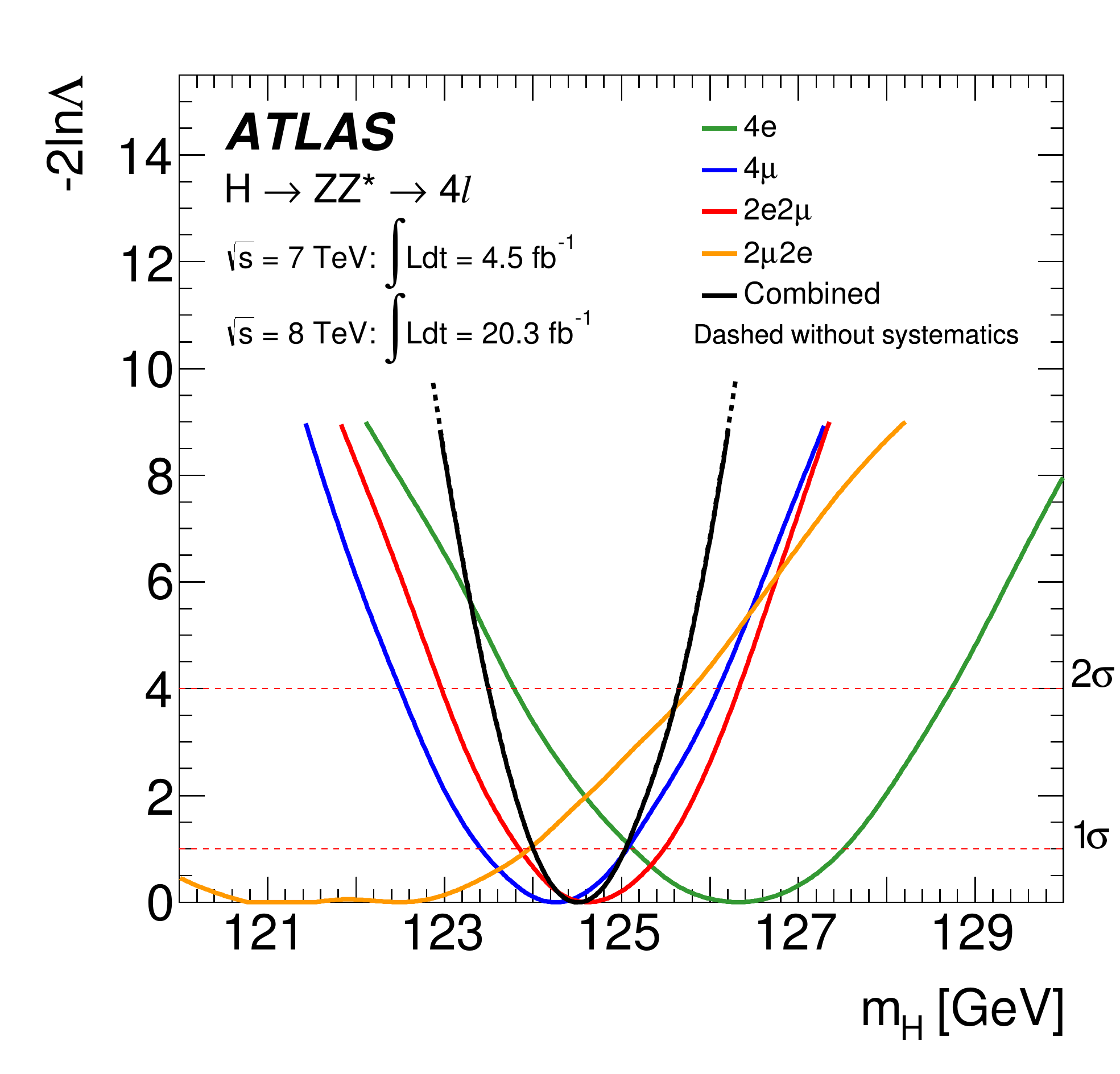}
  \caption{The profile likelihood as a function of $m_{H}$ for the combination of all
    \hzzllll\ channels and for the individual channels for the combined $7$~TeV and
    $8$~TeV data samples.  The combined result is shown both with (solid line) and without
    (dashed line) systematic uncertainties, and the two results are almost indistinguishable.
     \label{fig:like4lscan}}
\end{figure}

Using the per-event-error method a direct limit on the total width of the Higgs boson of
$\Gamma_{H}<$ 2.6 GeV at 95\% CL is obtained. The expected limit is $\Gamma_{H}<$ 6.2 GeV at 95\% CL
for a signal at the SM rate and $\Gamma_{H}<$ 3.5 GeV at 95\% CL for the observed signal rate.  The
difference between the observed and expected results arises from the higher signal strength observed
in the data, as well as from the measured $m_{4\ell}$, $O_{BDT_{\ZZbkg}}$ and mass resolution values
for the selected candidate events. These limits are estimated under the asymptotic assumption,
described in Section~\ref{sec:stat_and_sys}, and a cross-check with Monte Carlo ensemble tests
provides consistent results.  The limit on the total width was cross-checked with a 2D fit using
signal templates parameterized as a function of the Higgs boson width and found to be in agreement.

\section{Statistical procedure and treatment of systematic uncertainties\label{sec:stat_and_sys}}
The statistical treatment of the data is described in
Refs.~\cite{paper2012prd,LHC-HCG,Moneta:2010pm,HistFactory,Verkerke:2003ir}.  Confidence intervals are based
on the profile likelihood ratio $\Lambda(\vec\alpha)$~\cite{Cowan:2010js}. The latter depends on one
or more parameters of interest $\vec\alpha$, such as the Higgs boson mass $m_H$ or production yields
normalized to the SM expectation $\mu$, as well as on the nuisance parameters $\vec\theta$:

\begin{equation}
  \Lambda(\vec\alpha) = \frac{L\big(\vec\alpha\,,\,\hat{\hat{\vec\theta}}(\vec\alpha)\big)}
                              {L(\hat{\vec\alpha},\hat{\vec\theta})\label{eq:LH}}
\end{equation}

The likelihood functions in the numerator and denominator of the above equation are built using sums
of signal and background PDFs in the discriminating variables, such as the $\gamma\gamma$ mass
spectra for the \hgg\ channel and the $m_{4\ell}$ and  $\mathrm{BDT}_{\ZZbkg}$ output distributions
for the \hZZllll\ channel.  
The PDFs are derived from simulation for the signal and from both
data and simulation for the background, as described in Secs.~\ref{sec:hsg1} and \ref{sec:hsg2}.
Likelihood fits to the observed data are carried out for the
parameters of interest. The vector $\hat{\vec\theta}$
denotes the unconditional maximum likelihood estimate of the parameter
values and $\hat{\hat{\vec\theta}}$
denotes the conditional maximum likelihood estimate for given fixed values of the parameters of
interest $\vec\alpha$.  Systematic uncertainties and their correlations~\cite{paper2012prd} are
modeled by introducing nuisance parameters $\vec\theta$ described by likelihood functions associated
with the estimate of the corresponding effect. The choice of the parameters of interest depends on
the test under consideration, with the remaining parameters treated as nuisance parameters, {\it
i.e.} set to the values that maximize the likelihood function (``profiled'') for the given fixed
values of the parameters of interest.

For the combined mass measurement, hypothesized values of $m_H$ are tested using the profile
likelihood ratio defined in terms of $m_H$ and treating $\mu_{\gamma\gamma}(m_H)$ and
$\mu_{4\ell}(m_H)$ as independent nuisance parameters, so as to make no assumptions about the SM Higgs
couplings:
\begin{equation}\label{eq:teststatMh}
  \Lambda(m_H) = \frac{L\big(m_H\,,\,\hat{\hat{\mu}}_{\gamma\gamma}(m_H)\,,\,\hat{\hat{\mu}}_{4\ell}(m_H)\,,\,\hat{\hat{\vec\theta}}(m_H)\big)} {L(\hat{m}_H,\hat{\mu}_{\gamma\gamma},\hat{\mu}_{4\ell},\hat{\vec\theta})} \;.
\end{equation}

The leading source of systematic uncertainty on the mass measurement
comes from the energy and momentum scale uncertainties on the main physics objects used in the two analyses, namely photons for
the \hgg\ and muons and electrons for the \hZZllll\ final state. They are detailed in
Secs.~\ref{sec:egamma_sys} and ~\ref{sec:muon_sys}.  The correlation between the two measurements
stems from common systematic uncertainties and is modeled in the combination by correlating the
corresponding nuisance parameters. For the mass measurement this correlation comes mainly from the
uncertainty on the energy scale calibration with $Z \to e^{+} e^{-}$ events, which affects both the
electron and photon energy scale uncertainties. This source of
uncertainty is greatly reduced with respect to the previous
publication and has a small impact on the total mass uncertainty for
both channels. For this reason, the correlation between the two
measurements is now almost negligible.

To directly quantify the level of consistency between the measurements of $m_{H}^{\gamma\gamma}$ and
$m_H^{4\ell}$, the profile likelihood used for the mass combination is parameterized as a function of
the difference in measured mass values $\Delta m_{H} = m_{H}^{\gamma\gamma} - m_H^{4\ell}$, with the common mass $m_H$
profiled in the fit.
Specifically, the observable $m_H^{4\ell}$ is fit to the parameter $m_H$ while the observable
$m_{H}^{\gamma\gamma}$ is fit to the parameter $m_H + \Delta m_{H}$.  The two measurements are
compatible if the fitted value of $\Delta m_{H}$ is compatible with zero. The original model used to
combine the two measurements is recovered by fixing the parameter
$\Delta m_{H}$ to zero.

The signal strengths $\mu_{\gamma\gamma}$ and $\mu_{4\ell}$ are treated as independent nuisance
parameters in this approach, as is the common mass $m_H$.  The variation of $-2 \ln \Lambda(\Delta
m_H)$ between its minimum and the $\Delta m_{H} = 0$ point is used as an estimate of the
compatibility of the two masses, with all other fit parameters profiled to the data. This result
relies on the assumption that the statistical observable $-2\ln\Lambda$ behaves as a $\chi^2$
distribution with one degree of freedom, referred to as the asymptotic assumption. This
result is also cross-checked with Monte Carlo ensemble tests that do not rely on this assumption.
All sources of energy and momentum scale systematic uncertainty are treated assuming Gaussian PDFs.

\section{Combined mass measurement\label{sec:results}}
The measured masses from the \hgg\ and \hZZllll\ channels reported in Secs.~\ref{sec:hsg1}
and~\ref{sec:hsg2} are combined following the method described in Sec.~\ref{sec:stat_and_sys}. For
the \hZZllll\ channel the 2D method discussed in Sec.~\ref{sec:hsg2} is used.  The combined mass
measurement is:
\begin{equation}
\begin{aligned}
m_{H} &= 125.36 \pm 0.37 \,\mathrm{(stat)} \pm 0.18 \,\mathrm{(syst)} \gev \\  &= 125.36 \pm 0.41 \gev
\label{eq:mH}
\end{aligned}
\end{equation}
where the first error represents the statistical uncertainty and the second the systematic
uncertainty.  The statistical component is determined by repeating the likelihood scan with all
nuisance parameters related to systematic uncertainty fixed to their best fit value.  The systematic
component is then derived by subtracting in quadrature the statistical one from the total error.
The $-2\ln\Lambda$ value as a function of $m_{H}$ for the individual \hgg\ and \hZZllll\ channels
and their combination is shown in Fig.~\ref{fig:MassComb}.
\begin{figure}[tbhp!]
  \centering
  \includegraphics[width=\singlePlotSize]{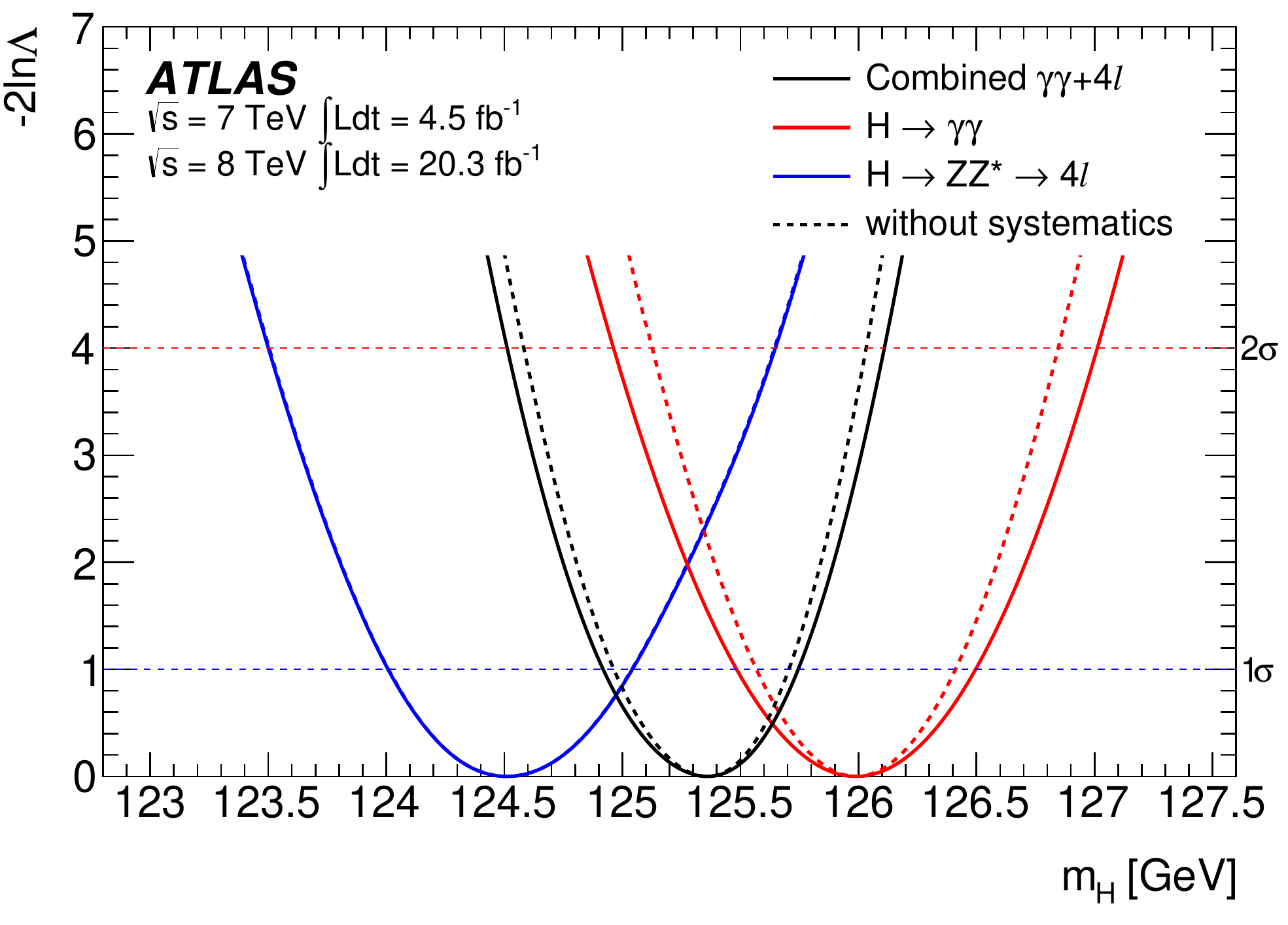}
  \caption{Value of $-2 \ln \Lambda$ as a function of $m_{H}$ for the individual \hgg\ and
    \hZZllll\ channels and their combination, where the signal strengths $\mu_{\gamma\gamma}$ and
    $\mu_{4\ell}$ are allowed to vary independently.  The dashed lines show the statistical
    component of the mass measurements.  For the \hZZllll\ channel, this is indistinguishable from
    the solid line that includes the systematic uncertainties.  }
   \label{fig:MassComb}
\end{figure}

With respect to the previously published value~\cite{ATLAScouplings} of $m_{H} = 125.49 \pm
0.24 \,\mathrm{(stat)} ^{+0.50} _{-0.58} \,\mathrm{(syst)} \gev $, the observed statistical error
has increased.  This is due to the increase of the observed statistical error in the \hgg\ channel
as discussed in Sec.~\ref{sec:hsg1_res}. The systematic uncertainty is significantly reduced thanks
to the improvements in the calibration of the photons and electrons and the reduction in the
uncertainty on the muon momentum scale, as detailed in Secs.~\ref{sec:egamma_sys}
and \ref{sec:muon_sys} respectively.

In order to check that the fitted signal yield is not significantly correlated with the measured
mass, the profile likelihood ratio as a function of both \mh\ and the normalized signal yield $S$,
$\Lambda(S,m_H)$ is used.  The normalized signal yield is defined as $S = \sigma/\sigma_{\rm
SM}(\mh{=}125.36\GeV)$.  It is similar to the signal strength $\mu = \sigma/\sigma_{\rm SM}(\mh)$,
except the \mh-dependence of the expected SM cross-sections and branching ratios that enter into the
denominator, principally for the \hZZllll\ channel, is removed by fixing \mh\ to the combined
best-fit mass.
Asymptotically, the test statistic $-2\ln\Lambda(S,m_H)$ is distributed as a $\chi^2$ distribution
with two degrees of freedom.  The resulting 68\% and 95\% CL contours are shown in
Fig.~\ref{fig:mass_2D_no_mh_dep}.  No significant correlation between the two fitted variables is
observed, confirming the model-independence of the mass measurement described in this paper.

\begin{figure}[tbhp!]
  \centering
  \includegraphics[width=\singlePlotSize]{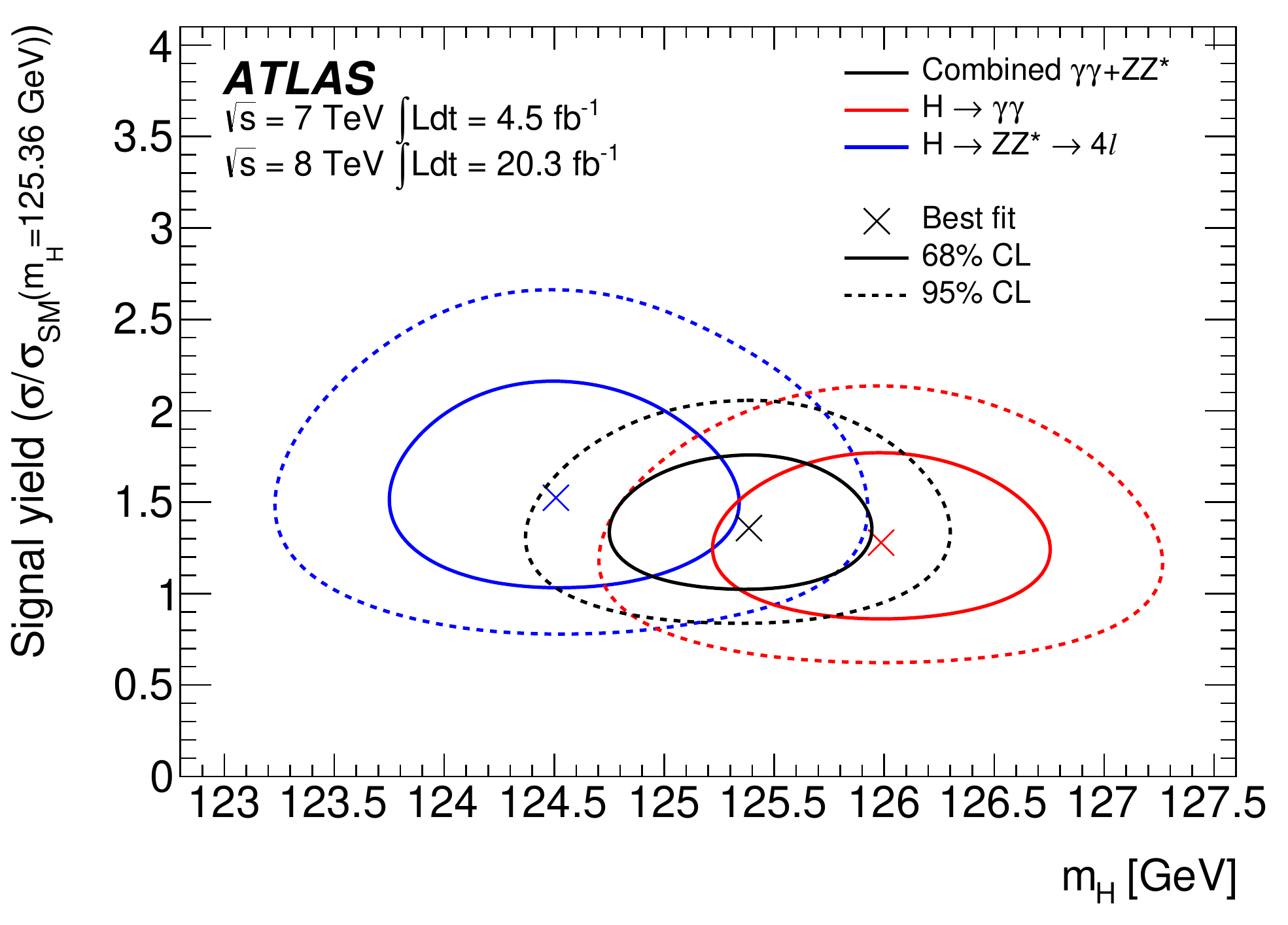}
  \caption{Likelihood contours $-2\ln\Lambda(S,m_H)$ as a function of the normalized signal yield $S
    = \sigma/\sigma_{\rm SM}(\mh{=}125.36\GeV)$ and \mh\ for the \hgg\ and \hZZllll\ channels
    and their combination, including all systematic uncertainties.
    For the combined contour, a common normalised signal yield $S$ is used.
    The markers indicate the maximum likelihood estimates in the corresponding channels.
      \label{fig:mass_2D_no_mh_dep}
  }
\end{figure}

As a cross-check, the mass combination was repeated fixing the values of the two signal
strengths to the SM expectation $\mu=1$.  The mass measurement only changes by 80 MeV, demonstrating that
the combined mass measurement is quite insensitive to the fitted values of the individual channel
signal strengths.

The contributions of the main sources of systematic uncertainty to the
combined mass measurement are shown in Table~\ref{tab:combined_sys}.
In the mass measurement fit, the post-fit values of the most relevant nuisance parameters, which are
related to the photon energy scale, do not show significant deviations from their pre-fit input values.

\begin{table*}
\center
\caption{Principal systematic uncertainties on the combined mass. Each uncertainty is
determined from the change in the 68\% CL range for $m_H$ when the corresponding
nuisance parameter is removed (fixed to its best fit value), and is calculated by subtracting this
reduced uncertainty from the original uncertainty in quadrature. \label{tab:combined_sys}}
\vspace{0.1cm}
\begin{tabular}{|l|r|}\hline
Systematic & Uncertainty on $m_H$ [MeV] \\
\hline
LAr syst on material before presampler (barrel) &  70 \\   
LAr syst on material after presampler (barrel)  &  20 \\   
LAr cell non-linearity (layer 2)                &  60 \\   
LAr cell non-linearity (layer 1)                &  30 \\   
LAr layer calibration (barrel)                  &  50 \\   
Lateral shower shape (conv)                     &  50 \\   
Lateral shower shape (unconv)                   &  40 \\   
Presampler energy scale (barrel)                &  20 \\   
ID material model ($|\eta| < 1.1$)              &  50 \\   
\hgg\ background model (unconv rest low $\ptt$) &  40 \\   
$Z\rightarrow ee$ calibration                   &  50 \\   
Primary vertex effect on mass scale             &  20 \\   
Muon momentum scale                             &  10 \\   
\hline
Remaining systematic uncertainties              &  70 \\
\hline\hline
Total                                           & 180 \\
\hline
\end{tabular}
\end{table*}

In order to assess the compatibility of the mass measurements from the two channels a dedicated test statistic
that takes into account correlations between the two measurements is used, as described in Sec.~\ref{sec:stat_and_sys}.
A value of
\begin{equation}
\begin{aligned}
\Delta m_{H} &= 1.47 \pm 0.67 \,\mathrm{(stat)} \pm 0.28 \,\mathrm{(syst)} \gev \\  &= 1.47 \pm 0.72 \gev
\label{eq:DeltaM}
\end{aligned}
\end{equation}
is derived. From the value of $-2\ln\Lambda$ at $\Delta m_{H}=0$, a compatibility of 4.8\%,
equivalent to $1.98\sigma$, is estimated under the asymptotic assumption.  This probability was
cross-checked using Monte Carlo ensemble tests.  With this approach a compatibility of 4.9\% is
obtained, corresponding to $1.97\sigma$.

As an additional cross-check, some of the systematic uncertainties related to the photon energy
scale, namely the inner detector material uncertainty and the uncertainty in the modeling of the
photon lateral leakage, were modeled using a ``box-like'' PDF defined as a double Fermi--Dirac
function.  This choice is compatible with the fact that for these uncertainties the data does not
suggest a preferred value within the systematic error range. In this case the compatibility between
the two masses increases to 7.5\%, equivalent to $1.8\sigma$.  The compatibility between the two
measurements increases to 11\% ($1.6 \sigma$) if the two signal strengths are set to the SM value of
one, instead of being treated as free parameters.

With respect to the value published in Ref.~\cite{ATLAScouplings}, 
the compatibility between the measurements from the individual channels has changed from $2.5\sigma$ to $2.0\sigma$.

\section{Conclusions\label{sec:conclusions}}
An improved measurement of the mass of the Higgs boson has been derived from a combined fit to the
invariant mass spectra of the decay channels \hgg\ and \hZZllll . These measurements are based on
the $pp$ collision data sample recorded by the ATLAS experiment at the CERN Large Hadron Collider at
center-of-mass energies of $\sqrt{s}$=7 TeV and $\sqrt{s}$=8~TeV, corresponding to an integrated
luminosity of 25\,\ifb. 
As shown in Table~\ref{tab:massSummary}, the measured values of the Higgs boson mass 
for the \hgg\ and \hZZllll\ channels are $125.98 \pm 0.42 \,\mathrm{(stat)} \pm 0.28 \,\mathrm{(syst)} \gev$ 
and $124.51 \pm 0.52 \,\mathrm{(stat)} \pm 0.06 \,\mathrm{(syst)} \gev$ respectively.
The compatibility between the mass measurements from the two individual channels is at the level 
of 2.0$\sigma$ corresponding to a probability of 4.8\%.  

From the combination of these two channels, the value of 
$m_{H} = 125.36 \pm 0.37 \,\mathrm{(stat)} \pm 0.18 \,\mathrm{(syst)} \gev$  is obtained.
These results are based on improved calibrations for photons, electrons and
muons and on improved analysis techniques with respect to Ref.~\cite{ATLAScouplings}, 
and supersede the previous results.

\begin{table*}[hptb!]
\centering
\caption{Summary of Higgs boson mass measurements. \label{tab:massSummary}}
\vspace{0.1cm}
\newcommand{\histrut}{\rule[-1.1ex]{0pt}{3.6ex}}
\begin{tabular}{|l|c|}\hline
Channel & Mass measurement [GeV] \\
\hline\hline
\histrut \hgg     & $125.98 \pm 0.42 \,\mathrm{(stat)} \pm 0.28 \,\mathrm{(syst)}  = 125.98 \pm 0.50 $ \\
\histrut \hZZllll & $124.51 \pm 0.52 \,\mathrm{(stat)} \pm 0.06 \,\mathrm{(syst)}  = 124.51 \pm 0.52 $ \\
\hline
\histrut Combined & $125.36 \pm 0.37 \,\mathrm{(stat)} \pm 0.18 \,\mathrm{(syst)}  = 125.36 \pm 0.41 $ \\
\hline
\end{tabular}
\end{table*}

Upper limits on the total width of the Higgs boson are derived from fits to the mass spectra of
the \hgg\ and \hZZllll\ decay channels, under the assumption that there is no interference with
background processes. In the \hgg\ channel, a 95\% CL limit of 5.0 (6.2) GeV is observed
(expected). In the \hZZllll\ channel, a 95\% CL limit of 2.6 (6.2) GeV is observed (expected).


\section{Acknowledgements}

We thank CERN for the very successful operation of the LHC, as well as the
support staff from our institutions without whom ATLAS could not be
operated efficiently.

We acknowledge the support of ANPCyT, Argentina; YerPhI, Armenia; ARC,
Australia; BMWF and FWF, Austria; ANAS, Azerbaijan; SSTC, Belarus; CNPq and FAPESP,
Brazil; NSERC, NRC and CFI, Canada; CERN; CONICYT, Chile; CAS, MOST and NSFC,
China; COLCIENCIAS, Colombia; MSMT CR, MPO CR and VSC CR, Czech Republic;
DNRF, DNSRC and Lundbeck Foundation, Denmark; EPLANET, ERC and NSRF, European Union;
IN2P3-CNRS, CEA-DSM/IRFU, France; GNSF, Georgia; BMBF, DFG, HGF, MPG and AvH
Foundation, Germany; GSRT and NSRF, Greece; ISF, MINERVA, GIF, I-CORE and Benoziyo Center,
Israel; INFN, Italy; MEXT and JSPS, Japan; CNRST, Morocco; FOM and NWO,
Netherlands; BRF and RCN, Norway; MNiSW and NCN, Poland; GRICES and FCT, Portugal; MNE/IFA, Romania; MES of Russia and ROSATOM, Russian Federation; JINR; MSTD,
Serbia; MSSR, Slovakia; ARRS and MIZ\v{S}, Slovenia; DST/NRF, South Africa;
MINECO, Spain; SRC and Wallenberg Foundation, Sweden; SER, SNSF and Cantons of
Bern and Geneva, Switzerland; NSC, Taiwan; TAEK, Turkey; STFC, the Royal
Society and Leverhulme Trust, United Kingdom; DOE and NSF, United States of
America.

The crucial computing support from all WLCG partners is acknowledged
gratefully, in particular from CERN and the ATLAS Tier-1 facilities at
TRIUMF (Canada), NDGF (Denmark, Norway, Sweden), CC-IN2P3 (France),
KIT/GridKA (Germany), INFN-CNAF (Italy), NL-T1 (Netherlands), PIC (Spain),
ASGC (Taiwan), RAL (UK) and BNL (USA) and in the Tier-2 facilities
worldwide.


\bibliographystyle{atlasBibStyleWoTitle} 
\bibliography{main} 



\onecolumn
\clearpage
\begin{flushleft}
{\Large The ATLAS Collaboration}

\bigskip

G.~Aad$^{\rm 84}$,
B.~Abbott$^{\rm 112}$,
J.~Abdallah$^{\rm 152}$,
S.~Abdel~Khalek$^{\rm 116}$,
O.~Abdinov$^{\rm 11}$,
R.~Aben$^{\rm 106}$,
B.~Abi$^{\rm 113}$,
S.H.~Abidi$^{\rm 159}$,
M.~Abolins$^{\rm 89}$,
O.S.~AbouZeid$^{\rm 159}$
% ...
\bigskip
\\
% ...
$^{11}$ Institute of Physics, Azerbaijan Academy of Sciences, Baku, Azerbaijan\\
% ...
$^{84}$ CPPM, Aix-Marseille Universit{\'e} and CNRS/IN2P3, Marseille, France\\
% ...
$^{89}$ Department of Physics and Astronomy, Michigan State University, East Lansing MI, United States of America\\
% ...
$^{106}$ Nikhef National Institute for Subatomic Physics and University of Amsterdam, Amsterdam, Netherlands\\
% ...
$^{112}$ Homer L. Dodge Department of Physics and Astronomy, University of Oklahoma, Norman OK, United States of America\\
$^{113}$ Department of Physics, Oklahoma State University, Stillwater OK, United States of America\\
% ...
$^{116}$ LAL, Universit{\'e} Paris-Sud and CNRS/IN2P3, Orsay, France\\
% ...
$^{152}$ Institute of Physics, Academia Sinica, Taipei, Taiwan\\
% ...
$^{159}$ Department of Physics, University of Toronto, Toronto ON, Canada\\
% ...
\end{flushleft}


\end{document}
